\documentclass[%
reprint,
showpacs,preprintnumbers,
amsmath,amssymb,
aps,
prb,
]{revtex4-1}
\usepackage{xcolor}
\usepackage{ulem}
\usepackage{mathtools, cuted}
\usepackage{amsmath}
\usepackage{lipsum}
\usepackage{lipsum, color}

\usepackage{graphicx}
\usepackage{dcolumn}
\usepackage{bm}
\usepackage{dcolumn}
\usepackage{dsfont}
\usepackage{bbm}
\usepackage{hyperref}
\usepackage[mathlines]{lineno}
\usepackage{natbib}
\begin{document}
\title{ Linear and nonlinear response for radiative heat transfer in many-body systems}
\author{A.~Naeimi} 
\author{M.~Nikbakht}
\email{mnik@znu.ac.ir}
\affiliation{Department of Physics, University of Zanjan, Zanjan 45371-38791,Iran.}
\date{\today}
\begin{abstract}
{
A  theory of temperature dynamics in many-body systems driven by time-dependent external sources is introduced. The formalism based on the combination of the perturbation theory and the fluctuational-electrodynamics approach in many-body systems. By using response theory, explicit formula for the temperature and phase shifts is derived and expressed in terms of the amplitude  and phase of external power sources.  Although the proposed method is highly efficient because it can skip the transient response, it is valid when external powers are weak.  As an illustration of this  theoretical framework, we have shown the dynamics of temperatures in one, two, and three degree of freedom systems driven by sine wave input powers. Finally, we highlighted some emergent phenomena arising from purely {\it dynamical} many-body effects, including amplification, attenuation, delaying or accelerating temperature responses.  This work could find important applications in the domain of dynamical thermal management at the nanoscale.}
\end{abstract}
\maketitle
\section{Introduction}\label{sec1}
Near-field radiative heat transfer at the nanoscale has been an interesting topic of many studies in last decade. When the separation distance between two objects is much smaller than the thermal wavelength $\lambda_T=\hbar c/k_BT$, heat transfer violates by the black-body limit predicted by Stefan-Boltzmann’s law. This phenomenon emerging as a consequence of the contribution of evanescent waves, known as {\it photon tunneling effect},  as recognized starting from the pioneering works of Cravalho {\it et al}.\cite{Cravalho}, Boehm {\it et al}.\cite{Boehm} and formulated by Polder and Van Hove\cite{PhysRevB.4.3303} based on the Rytov’s theory of fluctuational electrodynamics\cite{rytov1959theory}. 
Although the influence of objects characteristics on radiative heat transfer are of great importance and have been extensively studied\cite{PhysRevB.102.115417,moncadavilla2020normal,PhysRevB.99.134207,Choubdar,NIKBAKHT2018164,doi:10.1063/5.0018329}, much theoretical work has been done to improve our understanding of near-field and far-field radiative heat transfer in many-body systems\cite{PhysRevLett.107.114301,PhysRevB.95.205404,manybody,PhysRevB.96.125436,PhysRevB.95.235428,PhysRevB.95.125411,10.1088/1361-6633/abe52b,PhysRevB.100.205422,PhysRevB.96.155402}, and realizing and developing mechanisms to achieve magnitude and directional tunability of heat transfer between objects held at different temperatures\cite{doi:10.1063/1.4829618,doi:10.1063/1.4915138,doi:10.1021/acsnano.8b01645,PhysRevB.99.201406,PhysRevB.101.155428,PhysRevLett.117.134303,PhysRevB.97.094302,PhysRevB.102.085401,PhysRevApplied.13.034021}. Using these formalism, several studies have been conducted to investigate the thermal relaxation dynamics\cite{PhysRevB.88.104307,Nikbakht_2015,PhysRevLett.113.074301} and steady-state temperature profile\cite{PhysRevB.102.035433,QU2021107404} in few-body systems.
The management of heat transfer generally include active and passive approaches. The active heat transfer enhancement includes external parameters\cite{PhysRevB.101.085411,doi:10.1063/1.5145224,PhysRevApplied.13.054054,Dyakov_2015,PhysRevApplied.13.054054}, while the material compositions and geometric properties are belong to the passive heat transfer control\cite{PhysRevLett.112.044301,PhysRevB.101.241411}. The active control of heat transfer by means of external parameters motivated a tremendous effort focused on manipulating radiative heat transfer by means of external thermal power, electric field\cite{Volokitin2019,Volokitin_2020,Volokitin2020} and magnetic field\cite{PhysRevLett.118.173902,doi:10.1021/acsphotonics.7b01223,PhysRevB.97.205414,PhysRevB.92.125418}. Most of the recent activities in this field has dealt with  averaged or zero-frequency external parameters, including photonic thermal transistor\cite{doi:10.1063/1.4916730}, thermal refrigeration\cite{ZHOU2020106889,doi:10.1063/1.5018734}, thermal memories\cite{PhysRevLett.113.074301,PhysRevApplied.11.024004}, thermal logic gates\cite{PhysRevB.94.241401,kathmann2020scalable}, magneto thermo plasmonics   \cite{10.1117/1.JPE.9.032711,doi:10.1063/1.5093626}, and photon thermal Hall effect\cite{PhysRevLett.116.084301}.
Despite these research efforts, there have been a few studies on how the time-dependent parameters affect the thermal properties. Examples include radiative heat shuttling\cite{PhysRevLett.121.023903} and heat pumping\cite{PhysRevB.101.165435}.  
In practice it is very important to know how a system responds when it is driven by external sources at various frequencies. The linear response of a system, originally at equilibrium, to a time-dependent external source is proportional to the source, with proportionality coefficient obtained via response function \cite{ropke2013nonequilibrium}.
In this paper, we use the response theory to describe the dynamics of temperatures in N-body systems with arbitrary geometries. We focus on the steady-state temporal evolution of temperatures when system is driven by external time-dependent power sources.
To this aim, we use perturbation theory to derive closed-form analytical relation between the input powers and object's temperatures. After introducing the linear response matrix, we use the second order correction terms in temperatures to drive the second-order response matrix. We then
consider numerical applications on system of nanoparticles for one-degree of freedom to three degree of freedom cases, and show how the temperature of nanoparticles respond to an external sine powers. We find excellent agreement between the numerical simulations based on the response theory and the numerically evaluated exact response expression.  
Furthermore, time integration methods have low efficiency when solving for steady state responses since the transient responses must be solved first. The proposed method is highly efficient because it can skip the transient response and obtain the steady state response directly. As a results,  the formalism provides a method for fast and robust calculations of temperatures, and can be applied to a wide range of problems in radiative heat transfer, especially for high degree of freedom systems. We have also highlighted new features of many-body effects that appear in dynamic systems and can be used in active thermal management. The proposed formalism also makes it possible to study the thermal relaxation dynamics of objects as well as the calculation of temperature profile in the presence of constant input powers. 
The paper is structured as follows. In Sec.~\ref{sec2}, we present our physical system and introduce the formalism. Then, in Secs.~\ref{ss1} and \ref{ss2}, we formally derive the expressions of the linear response and the nonlinear response, respectively. 
In Sec.~\ref{sec3}, we give results of the model in few-particle system, and present our numerical applications to one- to three-degree of freedom system in section \ref{ss3} to \ref{ss5}. The influence of many-body interactions on the thermal response is investigated  in Sec.~\ref{ss6}. We
finally give some conclusive remarks in Sec.~\ref{sec4}.
\section{Theory and Model}\label{sec2}
We start by a system consist of N object exchanging energy via radiation. In the absence of any external source, the total power impinging on each object is shown by $\Phi_i$, and temperatures dynamic are given by 
\begin{equation}
\xi_i\frac{d T_i}{d t}=\Phi_i,~~(i=1,2,\cdots,N)
\label{eq1}
\end{equation}
where $\xi_i$ is the object heat capacity and $\Phi_i=\Phi_i^v+\Phi_i^c$ is the total internal power dissipated in object $i$, with  $\Phi_i^v$ being the total power impinging on object $i$ from all objects with varying temperature and $\Phi_i^c$ corresponds to the contribution of powers received from thermal bath and/or objects held at fixed temperatures. 
In the absence of any external power, one is usually interested in finding the fixed points ${\bf T}^{*}=[T_1^{*},T_2^{*},\cdots,T_N^{*}]$ of the dynamical system as those ${\bf\Phi}({{\bf T}^{*}})={\bf\Phi}^v({{\bf T}^{*}})+{\bf\Phi}^c=0$, where ${\bf\Phi}=[\Phi_1,\Phi_2,\cdots,\Phi_N]$. In the presence of {\it time dependent} external powers $F_i^e(t)$ (i.e., {\it driving term}), the dynamics of temperatures obey following relation
\begin{equation}
\xi_i\frac{d T_i}{d t}=\Phi_i+F_i^e(t)
\label{eq2}
\end{equation} 
No general solution to Eq.~(\ref{eq2}) for given applied external powers $F_i^e(t)$ is known. However, if applied powers are sufficiently weak compared to the internal powers, this equation can be solved by means of perturbation theory. For the sake of simplicity we assume that heat capacity of objects, $\xi_i$, are independent of temperature. The Taylor series expansion of the varying internal power ${\bf\Phi}_i^v$ in the neighborhood of ${\bf T}^{*}$ is
\begin{equation}
\begin{split}
\Phi_i^v({\bf T})&\simeq-\Phi_i^c\\
&+\sum_j\frac{\partial \Phi_i^v}{\partial T_j}(T_j-T_j^*)\\
&+\sum_j\frac{1}{2}\frac{\partial^2 \Phi_i^v}{\partial T_j^2}(T_j-T_j^*)^2.
\end{split}
\label{eq3}
\end{equation} 
Here we have used  ${\bf\Phi}^v({{\bf T}^{*}})=-{\bf\Phi}^c$. Since the total internal power is an additive function of dynamical variables $T_1, T_2, \cdots,T_N$, the off-diagonal elements of hessian matrix are zero and second derivative with respect to one variable appears in Eq.~(\ref{eq3}). The truncation of higher order terms relies on the assumption that input powers should be sufficiently weak compared to the internal powers. Therefore, by defining new variable $\textsf{T}_i=(T_i-T_i^*)$ as the temperature disturbances in the neighborhood of thermal equilibrium, Eq.~(\ref{eq2}) can be written as
\begin{equation}
\xi_i\dot{\textsf{T}}_i=\sum_j{\Phi'}^v_{ij}\textsf{T}_j
+\sum_{j}\frac{1}{2}{\Phi''}^v_{ij}\textsf{T}_j^2+F_i^e(t),
\label{eq4}
\end{equation} 
where we have introduced ${\Phi'}^v_{ij}=(\partial \Phi_i^v/\partial T_j)|_{{\bf \textsf{T}}=0}$ and ${\Phi''}^v_{ij}=(\partial^2 \Phi_i^v/\partial T_j^2)|_{{\bf \textsf{T}}=0}$. It is clear that ${\bf \textsf{T}}=0$ is a solution of this equation if $F=0$. For describing the dynamic characteristics of the system, we are using response theory by determining the response of the system to a sine wave input powers. If inputs are constant amplitude harmonic wave of fixed frequency
\begin{equation}
F_i^e(t)=F_{0i}e^{i\Omega t},
\label{eq5}
\end{equation} 
we are looking for the steady state temperature dynamics $\textsf{T}_i(t)$ upto second order of $F_{0i}$. To this end, we replace $F_{0i}$ in Eq.~(\ref{eq5}) by $\lambda F_{0i}$, where $\lambda$ is the perturbation strength and will be set equal to one at the end of calculations. Equation~(\ref{eq4}) then becomes
\begin{equation}
\xi_i\dot{\textsf{T}}_i=\sum_j{\Phi'}^v_{ij}\textsf{T}_j
+\sum_{j}\frac{1}{2}{\Phi''}^v_{ij}\textsf{T}_j^2+\lambda F_{0i}e^{i\Omega t}.
\label{eq6}
\end{equation} 
We now seek a solution to this equation in the form of power series expansion in the strength $\lambda$ of the perturbation, that is, a solution of the form
\begin{equation}
\textsf{T}_i=\textsf{T}_i^{(0)}+\lambda \textsf{T}_i^{(1)}+\lambda^2 \textsf{T}_i^{(2)}+\cdots
\label{eq7}
\end{equation} 
We restrict our calculation for the answers to the second order in $\lambda$. In order for Eq.~(\ref{eq7}) to be solution to Eq.~(\ref{eq6}) for any value of the perturbation parameter $\lambda$, it is required that dose terms in Eq.~(\ref{eq6}) proportional to $\lambda^0$, $\lambda^1$ and $\lambda^2$, each satisfy the equation separately. We observe that the terms proportional to $\lambda^0$, $\lambda^1$ and $\lambda^2$ lead respectively to the equations
\begin{subequations}
\begin{eqnarray}
&&\dot{\textsf{T}}_i^{(0)}=\xi_i^{-1}\left[\sum_j{\Phi'}^v_{ij}\textsf{T}_j^{(0)}+\sum_{j}\frac{1}{2}{\Phi''}^v_{ij}{\textsf{T}_j^{(0)}}^2\right],~~\label{eq8a}\\
&&\dot{\textsf{T}}_i^{(1)}=\xi_i^{-1}\sum_j{\Phi'}^v_{ij}\textsf{T}_j^{(1)}
~~\label{eq8b} \\
&&~~~~~~+\xi_i^{-1}\sum_{j}{\Phi''}^v_{ij}\textsf{T}_j^{(0)}\textsf{T}_j^{(1)}+\xi_i^{-1} F_{0i}e^{i\Omega t}\nonumber\\
&&\dot{\textsf{T}}_i^{(2)}=\xi_i^{-1}\sum_j{\Phi'}^v_{ij}\textsf{T}_j^{(2)}\label{eq8c}
\\
&&~~~~~~+\xi_i^{-1}\sum_{j}\frac{1}{2}{\Phi''}^v_{ij}{\textsf{T}_j^{(1)}}^2+\xi_i^{-1}\sum_{j}{\Phi''}^v_{ij}\textsf{T}_j^{(0)}\textsf{T}_j^{(2)}\nonumber.~~~~~~~
\end{eqnarray}
\end{subequations}
It is clear from Eq.~(\ref{eq8a}) that the lowest-order contribution ${\textsf{T}}^{(0)}$ is governed by the fixed points of Eq.~(\ref{eq4}) in the absence of external powers, setting $\dot{{\textsf{T}}}^{(0)}=0$ the trivial solution is ${\textsf{T}}^{(0)}=0$.
\subsection{Linear response}\label{ss1}
To evaluate the first-order correction term ${\textsf{T}}^{(1)}(t)$, we set ${\textsf{T}}^{(0)}=0$ in Eq.~(\ref{eq8b}) and seek a solution of the form 
 \begin{equation}
{\textsf{T}}_i^{(1)}(t)=\hat{\textsf{T}}_i^{(1)}(\Omega)e^{i\Omega t}.
\label{eq9}
\end{equation} 
Substituting Eq.~(\ref{eq9}) into Eq.~(\ref{eq8b}) leads to
\begin{equation}
{\hat{\bf {\textsf{T}}}}^{(1)}(\Omega)=\hat{\bf H}^{(1)}(\Omega) {\bf \mathcal{{F}}_0}, 
\label{eq10}
\end{equation} 
where 
\begin{equation}
{\hat{\bf {\textsf{T}}}}^{(1)}(\Omega)=
\begin{bmatrix} \hat{\textsf{T}}_1^{(1)}(\Omega)\\\hat{\textsf{T}}_2^{(1)}(\Omega)\\ \vdots \\\hat{\textsf{T}}_N^{(1)}(\Omega)\end{bmatrix},~~{\bf \mathcal{{F}}}_0=
\begin{bmatrix} \mathcal{F}_{01}\\\ \mathcal{F}_{02}\\ \vdots \\\mathcal{F}_{0N}\end{bmatrix},
\label{eq11}
\end{equation}
with scaled external power amplitude $\mathcal{F}_{0j}=\xi_j^{-1}F_{0j}$. Moreover, the linear response matrix is defined as
\begin{equation}
\hat{\bf H}^{(1)}(\Omega)=
\begin{bmatrix}
i\Omega-{\phi'}^v_{11} & -{\phi'}^v_{12} & \cdots & -{\phi'}^v_{1N}\\ -{\phi'}^v_{21} & i\Omega-{\phi'}^v_{22}& \cdots & -{\phi'}^v_{2N}\\ \vdots& \vdots & \ddots & \vdots\\ -{\phi'}^v_{N1} & -{\phi'}^v_{N2}& \cdots &i\Omega-{\phi'}^v_{NN}
\end{bmatrix}^{-1},
\label{eq12}
\end{equation}
where ${\phi'}_{ij}^v=\xi_i^{-1}{\Phi'}_{ij}^v$. The complex frequency response matrix $\small\hat{\bf H}^{(1)}(\Omega)$ contains information on all steady-state responses to harmonic excitation at different frequencies and amplitudes. The component $H^{(1)}_{ij}$ represents the contribution of the external power applied to object $j$ on the temperature dynamics of the object $i$. Using complex exponential notation, we can now write the corresponding harmonic linear response of object $i$ for harmonic input powers  as
\begin{equation}
{\textsf{T}}_i^{(1)}(t)=\sum_j \left[H^{(1)}_{ij}(\Omega)\mathcal{F}_{0j}e^{i\Omega t}\right]
\label{eq13}
\end{equation}
Finally, for $\sin$ wave input powers $F_j^e(t)=F_{0j}\sin(\Omega t)$, we have 
\begin{equation}
{\textsf{T}}_i^{(1)}(t)=\operatorname{\mathbb{I}m}\sum_j \left[H^{(1)}_{ij}(\Omega)\mathcal{F}_{0j}e^{i\Omega t}\right].
\label{eq14}
\end{equation}
Using the complex notation $H^{(1)}_{ij}={H'}^{(1)}_{ij}-i{H''}^{(1)}_{ij}$, it is straightforward to show that 
\begin{equation}
{\textsf{T}}_i^{(1)}(t)=\sum_j \tau_{j}^{(1)}\sin(\Omega t-\varphi_{ij}^{(1)}),
\label{eq15}
\end{equation}
where $ \tau_{j}^{(1)}=|H^{(1)}_{ij}|\mathcal{F}_{0j}$ and $\varphi_{ij}^{(1)}=\tan^{-1}({H''}^{(1)}_{ij}/{H'}^{(1)}_{ij})$.  In writing Eqs.~(\ref{eq13}) and (\ref{eq14}) in the forms shown, we have also assumed that the temperature  at time $t$ depends on the instantaneous value of the input powers strength as well as their frequency. Generally, the magnitude of the response function $|H^{(1)}_{ij}|$ gives the ratio of the amplitudes of input power and temperature response. Moreover, $\varphi_{ij}^{(1)}$ is the phase shift in the temperature response ${\textsf{T}}_i^{(1)}(t)$ with respect to the external thermal power impinging on $j$-th object. It should be notices that the power amplitudes $F_{0i}$ in Eq.~(\ref{eq5}) and so $\mathcal{F}_{0j}$ need not necessarily be real, as an example, for phase-shifted input powers of the form  $F_j^e(t)=F_{0j}\sin(\Omega t+\delta_j)$ we have ${\textsf{T}}_i^{(1)}(t)=\sum_j \tau_{j}^{(1)}\sin(\Omega t+\delta_j-\varphi_{ij}^{(1)})$.
We can deduce at once some of the properties of ${\textsf{T}}_i^{(1)}(t)$. As an example one can show that the time average of the first order correction to temperatures for harmonic input powers reduces to $\langle {\textsf{T}}_i^{(1)}\rangle=0$ or $\langle T_i\rangle=T_i^*$. It is also helpful to note that in the special case of constant external powers, we may set $\Omega=0$ in Eq~.(\ref{eq13}), and corresponding first order correction term to the steady state temperatures is given by ${\textsf{T}}_i^{(1)}=\sum_j H^{(1)}_{ij}(0)\mathcal{F}_{0j}$, and consequently $T_i=T^*_i+\sum_j H^{(1)}_{ij}(0)\mathcal{F}_{0j}$. The result is also consistent with Eq.~(\ref{eq1}), since in the absence of any external powers (i. e., $\mathcal{F}_{0j}=0$) we have $T_i=T^*_i$.
It should be emphasizes that equation~(\ref{eq13}) is general, since for aperiodic external powers $F_j^e(t)$ with scaled Fourier transformation 
\begin{equation}
\hat{\mathcal F}_j(\Omega)=\frac{1}{2\pi}\int_{-\infty}^\infty  \xi_j^{-1}F^e_j(t)e^{-i\Omega t}dt,
\label{eq16}
\end{equation}
the corresponding linear response function would be given by
\begin{equation}
{\textsf{T}}_i^{(1)}(t)=\sum_j\int_{-\infty}^\infty H_{ij}^{(1)}(\Omega)\hat{\mathcal F}_j(\Omega)e^{i\Omega t}d\Omega.
\label{eq17}
\end{equation}
with inverse Fourier transformation
\begin{equation}
{\textsf{T}}_i^{(1)}(t)=\int_{-\infty}^\infty \hat{\textsf{T}}_i^{(1)}(\Omega)e^{i\Omega t}d\Omega.
\label{eq18}
\end{equation}
By comparing Eq.~(\ref{eq17}) and Eq.~(\ref{eq18}) we obtain a transformation formula in the frequency domain as
 \begin{equation}
\hat{\textsf{T}}_i^{(1)}(\Omega)=\sum_jH_{ij}^{(1)}(\Omega)\hat{\mathcal F}_j(\Omega),
\label{eq19}
\end{equation}
which is an important relation between the Fourier transformation of the first order correction of temperatures and external powers and the first order response matrix. It should be emphasizes that the response functions $\hat{\bf H}^{(1)}$ is symmetric  (i.e., $H_{ij}^{(1)}=H_{ji}^{(1)}$) for reciprocal structures  and  its real and imaginary parts  are connected via the  Kramers Kronig relations. Moreover, since $\hat{\bf H}^{(1)}(-\Omega)=\hat{\bf H}^{(1)*}(\Omega)$, it implies that ${H'}^{(1)}(\Omega)$ is an even function of $\Omega$ and ${H''}^{(1)}(\Omega)$ is an odd function of $\Omega$. 
It is also interesting to notice that in the absence of  external thermal powers Eq.~(\ref{eq8b}) reduces to a linear relation $\dot{\textsf{T}}^{(1)}=\left[\hat{\bf H}^{(1)}(0)\right]^{-1}{\textsf{T}}^{(1)}$ with solution that can be visualized as phase trajectories moving toward the fixed point ${\bf \textsf{T}}^{(1)}(t\to \infty)=({\bf T}-{\bf T^*})=0$. The theory of linear algebra allows us to write down the general solution as ${\bf T}^{(1)}(t)=\sum_i C_i\exp(\lambda_i t){\bf V_i}$ where $\lambda_i$ and $\bf V_i$ are the eigenvalues  and eigenvectors of the Jacobian matrix $\left[\hat{\bf H}^{(1)}(0)\right]^{-1}$, respectively. Moreover, coefficients $C_i$'s are given by initial condition for temperatures according to  ${\bf T}^{(1)}(0)=\sum_i C_i{\bf V_i}$.
\subsection{Nonlinear response}\label{ss2}
In order to calculate the nonlinear response to the external input powers,  we seek a solution of the form 
 \begin{equation}
{\textsf{T}}_i^{(2)}(t)=\hat{\textsf{T}}_i^{(2)}(\Omega)e^{2i\Omega t}.
\label{eq20}
\end{equation}  
To this end,  the expression for ${\textsf{T}}_i^{(1)}$ in Eq.~(\ref{eq10}) is squared and substituted into Eq.~(\ref{eq8c}) which is solved to obtain the second-order correction terms $\hat{\textsf{T}}_i^{(2)}(\Omega)$. Recalling that ${\textsf{T}}_i^{(0)}=0$, we obtain
\begin{equation}
{\hat{\textsf{T}}_i}^{(2)}(\Omega)=\sum_{j}H^{(2)}_{ij}(\Omega){\mathcal F}_{0j}, 
\label{eq21}
\end{equation} 
with
\begin{equation}
H^{(2)}_{ij}(\Omega)=\sum_{klm}\frac{{\phi''}_{kl}^v}{2}H^{(1)}_{ik}(2\Omega)H^{(1)}_{lm}(\Omega)H^{(1)}_{lj}(\Omega){\mathcal F}_{0m}.
\label{eq22}
\end{equation} 
where ${\phi''}_{kl}^v=\xi_k^{-1}{\Phi''}_{kl}^v$. 
From Eq.~(\ref{eq20}) and Eq.~(\ref{eq21}), we can now write the corresponding harmonic nonlinear response of object $i$ for harmonic input powers  as
\begin{equation}
{\textsf{T}}_i^{(2)}(t)=\sum_{j}\left[H^{(2)}_{ij}(\Omega)\mathcal{F}_{0j}e^{2i\Omega t}\right],
\label{eq23}
\end{equation}
which can also be expressed in terms of first-order temperature correction terms as
\begin{equation}
{\textsf{T}}_i^{(2)}(t)=\sum_{jk}\left[\frac{{\phi''}_{jk}^v}{2}H^{(1)}_{ij}(2\Omega){{\textsf{T}}_k^{(1)}}^2(t)e^{-2i\Omega t}\right].
\label{eq24}
\end{equation}
Finally, for phase-shifted constant amplitude sine wave input powers we have 
\begin{equation}
{\textsf{T}}_i^{(2)}(t)=\operatorname{\mathbb{I}m}\sum_{jk}\left[\frac{{\phi''}_{jk}^v}{2}H^{(1)}_{ij}(2\Omega){{\textsf{T}}_k^{(1)}}^2(t)e^{-2i(\Omega t+\delta_k)}\right] \\
\label{eq25}
\end{equation}
From Eq.~(\ref{eq22}), it is clear that the dependency of second-order frequency response function on $\Omega$, is nonlinear as in $\hat{\bf H}^{(1)}(\Omega)$.  Is is also very important to notice that $\hat{\bf H}^{(2)}(\Omega)$  depends not only on the frequency, as in the linear case, but also on the amplitude and phase of the input powers. This implies, we may observe a non-proportional change in the magnitude and phase of the second-order correction term with respect to a change of the external power amplitudes and phases. The direct consequence of this non-linearity is that while the time average of the first order correction term always vanishes as discussed earlier, however, in general $\langle T_i\rangle\neq T^*_i$ even for harmonic input powers. On the other hand, by taking into account the  contribution of second-order correction term we have $\langle T_i\rangle= T^*_i+\langle T^{(2)}_i\rangle$. We can estimate the validity of linear regime by noting that the linear and nonlinear contributions to the temperature dynamics given by Eq.~(\ref{eq4}) would be expected to become comparable when the thermal conductance $\Phi'_{ij}|_{T^*_i}$ is approximately equal to $\Phi''_{ij}|_{T^*_i}$. Hence, it is expected for the linear response to be more accurate at larger thermal equilibrium temperature $T_i^*$. We can also make use of Eqs.~(\ref{eq13}) and (\ref{eq23}) for response functions to estimate the contribution of these terms with respect to the frequency and strength of the input powers.  We observe a power law decrease of the form $|{H}^{(1)}|\simeq \Omega^{-1}$ and $|{H}^{(2)}|\simeq \Omega^{-3}$  at large frequencies (i. e., $\Omega\gtrsim \Phi'/\xi$) for linear and nonlinear response functions, respectively. This suggests that the  nonlinear term can be ignored at higher frequency input powers. On the other side, one can show from Eq.~(\ref{eq22}) that the condition for negligible nonlinear term reduces to $F_0\ll \frac{{\Phi'}^2}{\Phi''\xi}$, which  indicate that the linear response theory is more accurate for system of particles with smaller heat capacity. It should be added that relations (\ref{eq15}) and (\ref{eq25}) allow also the study of temporal behavior of temperatures  for sum- and difference-frequency generation. Without making here in detail such an analysis (this will be done in a future work) we
restrict ourselves to input powers at same frequencies. 
\section{Results and discussion}\label{sec3}
Let us now apply this theoretical framework to describe the temperature dynamics in a many-body systems. To this end, we consider a system composed of spherical nanoparticles exchange energy via radiation and placed in thermal baths at fixed temperature $T_b$. The total photonic power received by  $i^{th}$ particle is given by $\Phi_i=\int_0^\infty \frac{d\omega}{2\pi}{\mathcal T}_{ib}\Theta(\omega,T_b)+\sum_{j=1}^N\int_0^\infty \frac{d\omega}{2\pi}{\mathcal T}_{ij}\Theta(\omega,T_j)$, where $\Theta(\omega , T)=\hbar\omega/[\exp{(\hbar\omega/k_BT)}-1]$ is the mean energy of Planck oscillator at temperature $T$. Furthermore, ${\mathcal T}_{ij}(\omega)$ denotes the energy transmission coefficient between $i^{th}$ and $j^{th}$ particles and ${\mathcal T}_{ib}(\omega)$ stands for the energy transmission coefficient recieved from the surrounding thermal bath. These transmission coefficients are\cite{manybody,Choubdar,PhysRevB.88.104307}
\begin{equation}
{\mathcal T}_{ij}(\omega)=2\operatorname{\mathbb{I}m}{\mathbb Tr}[{\mathbb A}_{ij}\operatorname{\mathbb{I}m}{\chi}_j{\mathbb C}_{ij}^{\dagger}],
\label{eq26}
\end{equation}
and
\begin{equation}
{\mathcal T}_{ib}(\omega)=2\operatorname{\mathbb{I}m}{\mathbb Tr}[{\mathbb B}_{ij}\operatorname{\mathbb{I}m}{G}_{jj'}{\mathbb D}_{ij'}^{\dagger}],
\label{eq27}
\end{equation}
where $\chi$ is the susceptibility function, and ${\mathbb A}_{ij}$, ${\mathbb B}_{ij}$, ${\mathbb C}_{ij}$, ${\mathbb D}_{ij'}$ and $G_{jj'}$ are given in terms of Green's functions and particle polarizabilities in many-body system.
For concreteness, we will consider hexagonal Boron Nitride (hBN) nanoparticles, for which the dielectric permittivity is well described by a Drude-Lorentz model $\epsilon(\omega)=\epsilon_\infty(\omega_L^2-\omega^2-i\gamma\omega)/\omega_T^2-\omega^2-i\gamma\omega)$, where $\epsilon_\infty=4.9$, $\omega_L=3.03\times10^{14}$~rad/s, $\omega_T=2.57\times10^{14}$ and $\gamma=3.2\times10^{12}$~rad/s. In the following, we apply the developed formalism to analyse the steady-state in one, two and three dimensional dynamical systems. The calculated temperature dynamics based on the response theory are verified by comparison with the results calculated directly from the exact fluctuation electrodynamics theory. It should be emphasize here that in the following numerical examples we always consider the steady-state regime. 
\begin{figure}
\includegraphics[]{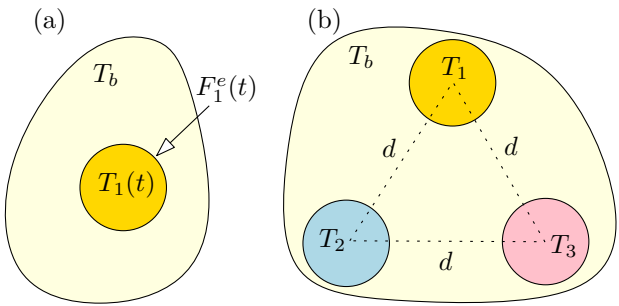}
\caption{ Geometry of the one-dimensional dynamical system. (a) One-body system: A nanoparticle with radius $R=100$~nm is placed in a thermal bath at temperature $T_b=300$~K and subjected to an external sine wave power input $F^e(t)$. (b) Three-body system: System consists of three equi-distant particles 1, 2 and 3 with $d=600$~nm and $R=100$~nm. System is placed in an external thermal bath at temperature $T_b=300$~K and particle 1 is subjected to an external sine wave power input $F^e(t)$. Furthermore, particles 2 and 3 are assumed to be perfectly coupled with two thermostats at temperature $T_2=300$~K and $T_3=350$~K, respectively.}
\label{Figure.1}
\end{figure}
\begin{figure}
\includegraphics[]{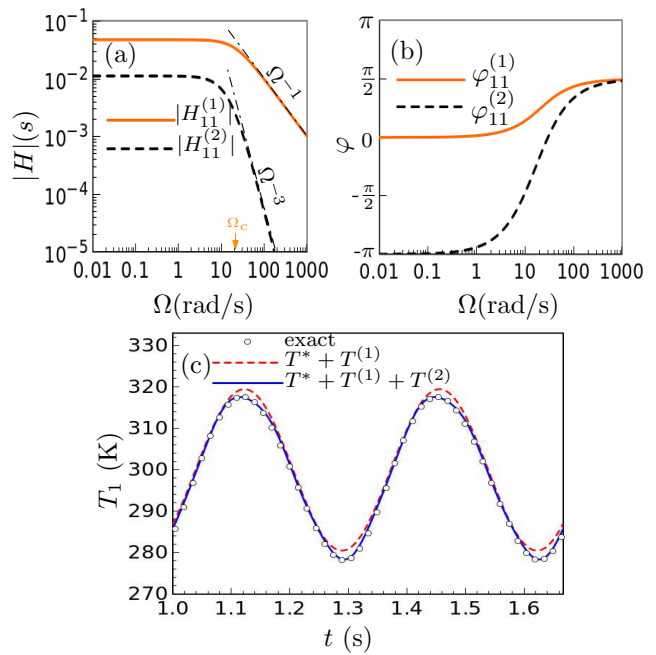}
\caption{  (a) Magnitude and (b) Phase for first-order and second order response functions of a single degree of a one-body system. (c) Temperature dynamics of a single nanoparticle in one-body system depicted in Fig.~(\ref{Figure.1}a), subjected to an external power $F^e(t)=F_1\sin(\Omega t)$ with $F_1=6\times 10^{-13}$~W and $\Omega=6\pi$~rad/s. The calculated result from the response theory up to first (second) order are shown in dash-red (solid-blue) lines. The result calculated from fluctuational electrodynamics with the use of Runge–Kutta 4 method is shown as "exact".}
\label{Figure.2}
\end{figure}
\subsection{One degree of freedom system}\label{ss3}
Let us consider the situation depicted in Fig.~(\ref{Figure.1}a), where a spherical nanoparticle with radius $R=100$~nm is placed in an external thermal bath with fixed temperature $T_b=300$~K. In the absence of an external power, the nanoparticle thermalizes to the bath temperature, i.e., $T^*=300$~K. Now, the particle is subjected to an external thermal source $F^e(t)=F_1\sin(\Omega t+\delta)$. 
The log-log shape of $|H|$ versus frequency is plotted in Fig.~(\ref{Figure.2}a).  The diagram  is practically the same as a low-pass filter with three sections. We observe a constant asymptotic behavior at low frequency, an inflexion point $\Omega_c\sim \Phi'/\xi$, and finally a power law decrease at high frequencies. The dependence of  $\varphi$ on frequency $\Omega$ is shown in Fig.~(\ref{Figure.2}b). We observe in Fig.~(\ref{Figure.2}a) and Fig.~(\ref{Figure.2}b), that at low frequencies $|{H}^{(1)}_{11}|\simeq \Omega_c^{-1}$ and $\varphi^{(1)}_{11}\sim 0$ which implies that we get the maximum response with same phase as the input power. At the same time, the magnitude of second order response function is very small compared to the first order term ($|{H}^{(2)}_{11}|\simeq 0.5F_1\phi''_{11}/{\phi'}^2_{11}$) and have a negative contribution in temperature since $\varphi^{(2)}_{11}<0$. However, for high frequency input powers we have $|{H}^{(1)}_{11}(\Omega)|\propto \Omega^{-1}$, $|{H}^{(2)}_{11}(\Omega)|\propto \Omega^{-3}$, and $\varphi_{11}\to \pi/2$. On the other hand, if we subject the particle to a high frequency oscillatory input power, there will be almost no time for particle to respond before the powers have switched direction, and so the frequency response  will converge to zero as $\Omega$ becomes very large. 
\begin{figure}
\includegraphics[]{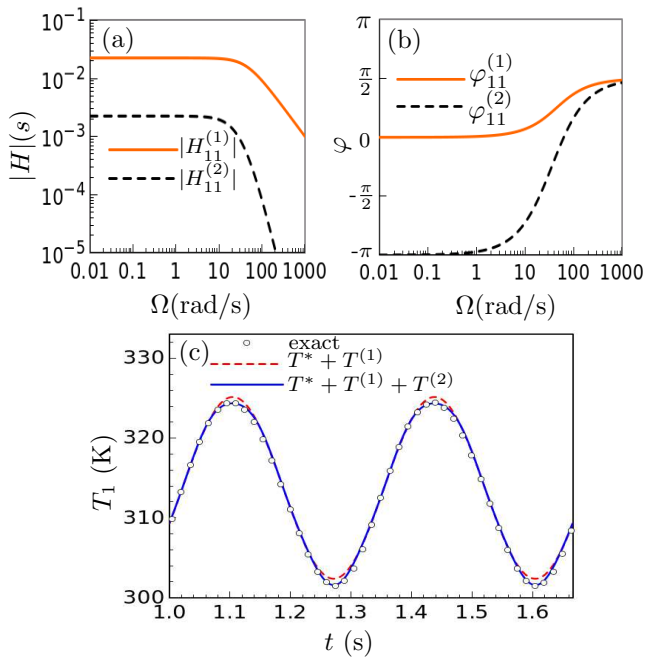}
\caption{  (a) Magnitude and (b) Phase for the first and second order response functions of a single degree of freedom in three-body system depicted in Fig.~(\ref{Figure.1}b). (c) Temperature dynamics of a single nanoparticle in three-body system. Particle 1 is subjected to an external power $F^e(t)=F_1\sin(\Omega t)$ with $F_1=6\times 10^{-13}$~W and $\Omega=6\pi$~rad/s, while particles 2 and 3 are connected to reservoirs at fixed temperatures $T_2=300$~K and $T_3=350$~K, respectively. The calculated result from the response theory up to first (second) order are shown in dash-red (solid-blue) lines. The result calculated from fluctuational electrodynamics with the use of Runge–Kutta 4 method is shown as "exact".}
\label{Figure.3}
\end{figure}
For the special case of $F_1=6\times 10^{-13}$~W, $\Omega=6\pi$~rad/s, and $\delta=0$ the temperature dynamic solved based on the developed response theory is shown in Fig.~(\ref{Figure.2}c). Results show that the temperature dynamics calculated from the first-order response theory agrees very well with the exact result with maximum relative error of $0.6\%$, which improved by second-order term and gives the maximum  relative error of $0.03\%$. 
We will again use the proposed theory as the basis for our discussion, but we will now include many-body effects by adding two nanoparticles to the first configuration.  As shown in
Fig.~(\ref{Figure.1}b), the system consists of three equi-distant nanoparticles with $d=600$~nm and radius $R=100$~nm, in which objects 2 and 3 are at fixed temperatures $T_2=300$~K and $T_3=350$~K, respectively.  In the presence of an external thermal bath at temperature $T_b=300$~K, the fixed point of $T_1$ in this configuration changes to $T_1^*=313.764$~K, and imposing external sine power wave to particle 1 (as in one-body case) results in an oscillatory behavior of $T_1(t)$. The frequency dependence of the  magnitude and phases of the response functions are shown in Fig.~(\ref{Figure.3}a) and Fig.~(\ref{Figure.3}b), respectively. It is clear that the magnitude of the response functions are smaller than that of one-body system, ascribed to the modification in response functions due to the presence of nanoparticles 2 and 3 (i.e., three-body effects). We also observe that the response phase shifts, shown in Fig.~(\ref{Figure.3}b), have similar trends as in one-body case. The temperature dynamic solved based on the developed response theory and the exact results are shown in Fig.~(\ref{Figure.3}c). Once again, we observe a good agreement between the temperature dynamics calculated from the response theory and that of exact method in the presence of many-body effects. 
\begin{figure}
\includegraphics[]{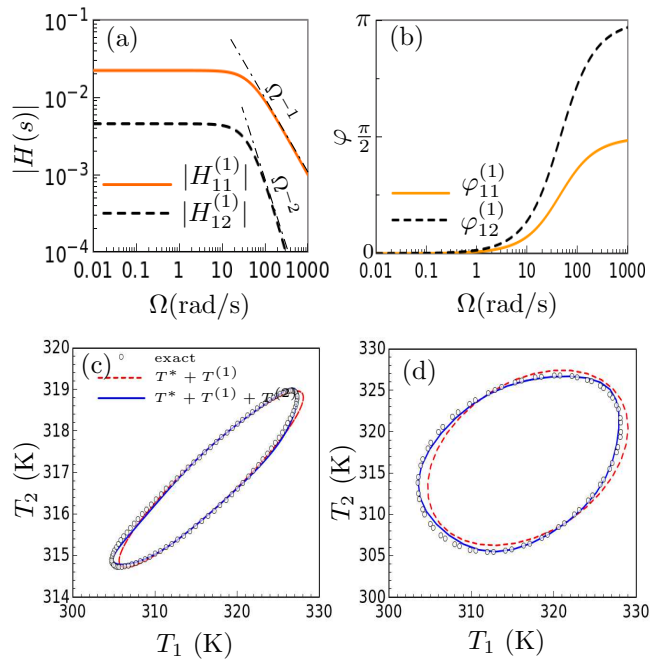}
\caption{(a) Magnitude and (b) Phase of the components of the first order response matrix for two-degree of freedom in three-body system depicted in Fig.~(\ref{Figure.1}b). Particles 1 and 2 are driven by external sine power and particle 3 is at constant temperature $T_3=350$~K. Steady-state phase trajectory of the system for (c) $F_1=6\times 10^{-13}$~W, $F_2=0$, and (d) $F_1=6\times 10^{-13}$~W, $F_2=6\times 10^{-13}$~W, $\delta_2=\pi/2$. The calculated result from the response theory up to first (second) order are shown in dash-red (solid-blue) lines. The result calculated from fluctuational electrodynamics with the use of Runge–Kutta 4 method is shown as "exact".}
\label{Figure.4}
\end{figure}
\subsection{Two degree of freedom system}\label{ss4}
In order to investigate the response theory for two-dimensional dynamical system, we have used a same configuration as in Fig.~(\ref{Figure.1}b). However, the temperature of the second object is not fixed any more and allowed to vary with time. The frequency response of the system would be a $2\times 2$ matrix, where in addition to the intrinsic symmetry of the first-order response matrix (i.e., $H^{(1)}_{12}=H^{(1)}_{21}$), the diagonal elements are also equal to one another  (i.e., $H^{(1)}_{11}=H^{(1)}_{22}$) due to the configuration symmetry. The dependence of the corresponding amplitudes and phases on the frequency are shown in Fig.~(\ref{Figure.4}a) and Fig.~(\ref{Figure.4}b), respectively. We observe in Fig.~(\ref{Figure.4}a), that the amplitude of the off-diagonal elements are smaller than the diagonal elements and  for high frequency input powers we have $|{H}^{(1)}_{12}(\Omega)|\propto \Omega^{-2}$. On the other side, while the power applied to each particle has an immediate effect on the temperature of the other particle (i.e., $\varphi^{(1)}_{12}= \varphi^{(1)}_{21}\to 0$) for small frequency input powers, there would be a phase lag of $\pi$ for high frequency input powers.
\begin{figure}
\includegraphics[]{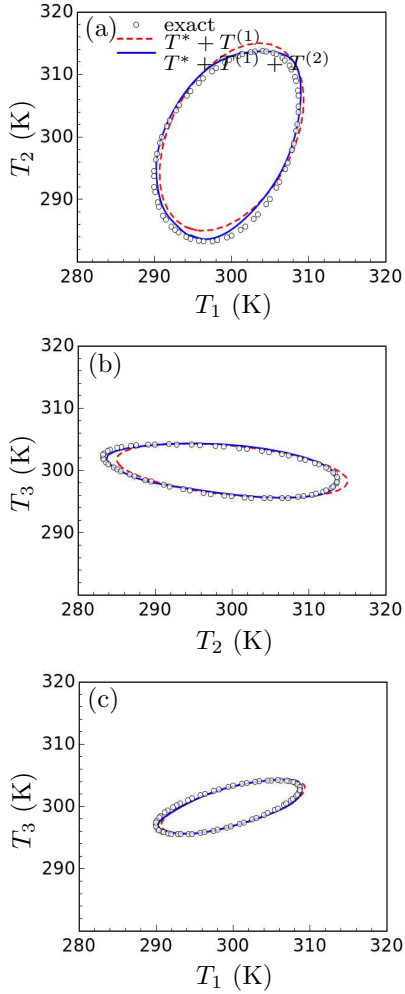}
\caption{  Steady-state phase trajectory of temperatures in three-degree of freedom system projected onto (a) $T_1-T_2$ plane, (b) $T_2-T_3$ plane and (c)  $T_1-T_3$ plane. System consists of three equi-distant particles  initially at thermal equilibrium at ambient temperature $T_b=300$~K and driven by external sine power waves with $F_1=4\times 10^{-13}$~W, $F_2=6\times 10^{-13}$~W, $F_3=3\times 10^{-13}$~W, $\delta_2=-\pi/2$ and  $\delta_3=\pi/3$. The calculated result from the response theory up to first (second) order are shown in dash-red (solid-blue) lines. The result calculated from fluctuational electrodynamics with the use of Runge–Kutta 4 method is shown as "exact".}
\label{Figure.5}
\end{figure}
Depending on the amplitude and phase of the input powers, the phase trajectory approaches an isolated periodic orbit (a limit cycle) after the decay of initial transients. In order to examine the theory, two cases are investigated in such a second degree of freedom system. Figure~(\ref{Figure.4}c) shows the results for input sine powers with $F_1=6\times 10^{-13}$~W, $\Omega=6\pi$~rad/s, and $F_2=0$. On the other side both objects are driven by sine wave power with $F_1=6\times 10^{-13}$~W, $F_2=6\times 10^{-13}$~W, $\Omega=6\pi$~rad/s, and $\delta_2=\pi/2$, see Fig.~ (\ref{Figure.4}d). For the former, we observe that the output of the linear response theory is very close to the exact result, and second-order response function has completely improved it. For the latter, constructive overlap of applied powers increased the non-linearity and linear response is not accurate. However, we observe a good agreement between the second-order corrected temperatures  with does by the exact method. 
\subsection{Three degree of freedom  system}\label{ss5}
Finally, we consider a three degree of freedom problem in a three-body system. We have used a same configuration as in Fig.~(\ref{Figure.1}b),  where three equi-distant nanoparticles with separation distance $d=600~$nm and radius $R=600~$nm are initially at thermal equilibrium at ambient temperature $T_b=300$~K. We assume that temperatures $T_1$, $T_2$ and $T_3$ are free to change and particles are driven by external sine powers with $F_1=4\times 10^{-13}$~W, $F_2=6\times 10^{-13}$~W,  $F_3=3\times 10^{-13}$~W, $\delta_2=-\pi/2$ and  $\delta_3=\pi/3$.  As expected on physical grounds, regardless of the initial conditions, the  phase trajectory of the system approach a limit cycle in three dimensional phase space $T_1-T_2-T_3$. The projections of the  steady-state limit cycle  onto two-dimensional planes $T_1-T_2$, $T_2-T_3$ and $T_1-T_3$ for both the exact and response methods are shown in Fig.~(\ref{Figure.5}). Once again we observe that the linear response method give a qualitatively correct picture of phase trajectory at the steady-state regime. As maintained earlier, the product of input power amplitudes like $F_1^2$, $F_2^2$, $F_3^2$, $F_1F_2$, $F_2F_3$ and $F_3F_1$ do in fact lead to a contribution to the nonlinear temperature dynamics. Since amplitude of the two input powers $F_1$ and $F_2$ are larger than $F_3$, we observe a larger deviation of results between the linear response method and that of the exact method in Fig.~(\ref{Figure.5}a). On the other side, the non-linearity decreased in Fig.~(\ref{Figure.5}b) for the dynamics of temperatures in $T_2-T_3$ subspace. Furthermore, since $F_1,F_3<F_2$, we observe that the linear response method seems to well predict the dynamics of temperatures in  $T_1-T_3$ subspace as shown in  Fig.~(\ref{Figure.5}c).
\begin{figure}
\includegraphics[]{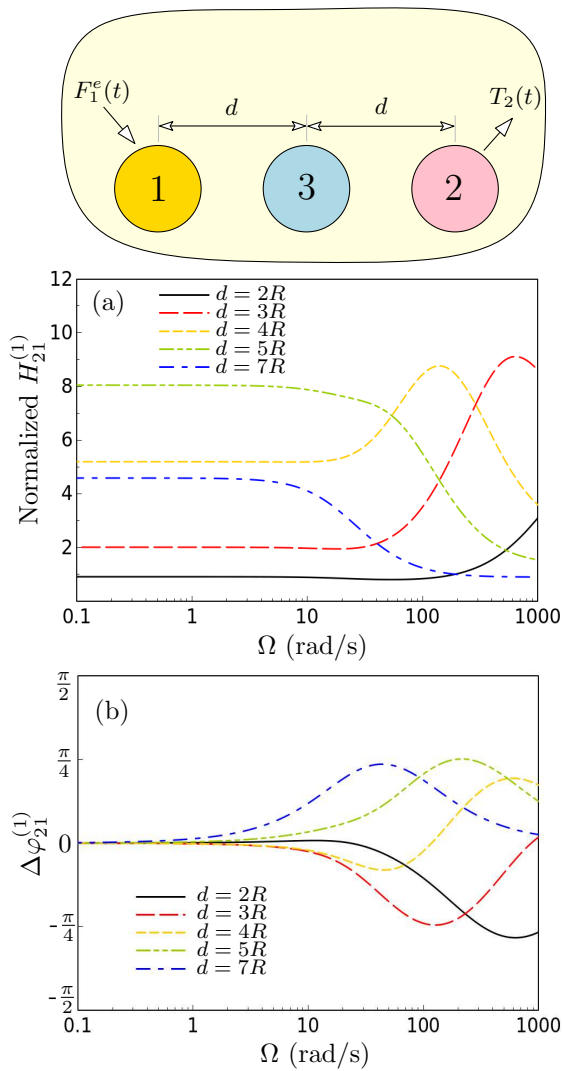}
\caption{ Dynamical many-body effects: (a) Linear response of particle 2 to the input power applied to particle 1 in a three-body system normalized to the two-body case ($|H_{21}^{(1)}|_{\textrm{3-body}}/|H_{21}^{(1)}|_{\textrm{2-body}}$), and (b) Phase lag difference $\Delta\varphi=\varphi_{21}^{(1)}|_{\textrm{3-body}}-\varphi_{21}^{(1)}|_{\textrm{2-body}}$ as a function of the input power frequency for various separation distances $d$.}
\label{Figure.6}
\end{figure}
\subsection{Many-body effects}\label{ss6}
 In order to explicitly reveal the many-body effects, we consider a system composed of three aligned nanoparticles with radius $R=100~$nm and labeled with indexes 1, 2 and 3 as shown in Fig.~\ref{Figure.6}.
The third particle is located between the two other particles with equal distances $d$. We assume that temperatures $T_1$, $T_2$ and $T_3$ are free to change and the entire system is initially at thermal equilibrium at ambient temperature $T_b=300$~K. It is clear from Eq.~(\ref{eq13}), that if we apply an external harmonic power $F_1^e(t)=F_1\sin(\Omega t)$ to particle 1, it affects the temperature of particle 2, which is $ T_2(t)\sim T_b+|H_{21}^{(1)}(\Omega)|\mathcal{F}_{1}\sin(\Omega t+\varphi_{21}^{(1)})$. Regardless of the value of the external power amplitude $F_1$, the contribution of this effect is proportional to the element $H_{21}(\Omega)$ of the response matrix, which depends on the existence and position of the third particle. To get an idea of this effect, the linear response of the temperature of particle 2 to an external power $F_1^e(t)$ is calculated in three-body system and normalized to the value in the absence of the third particles. Calculation are performed for several separation distances and results are shown in Fig.~(\ref{Figure.6}a). As can be seen in this figure, the thermal response mediated by the presence of the third particle can be larger than the value we could have in two-body system. Notice that the enhancement in response function depend not only on the frequency of the input power, but also on the properties defining the geometry such as separation distance in this case.  For slowly varying external power sources, the normalized response function depends only on the geometric parameter $d$, and can be engineered similar to the previously observed many-body effects in static limit \cite{PhysRevLett.107.114301}. On the other side, we observe that the many-body effects could be more pronounced in the dynamic cases where $\Omega\neq 0$. In particular, it can be seen from this figure that for a fixed geometrical parameter $d$, the normalized response function depends on the frequency of the input power. While this {\it dynamical} many-body effect is a decreasing function of frequency for large separation distances, it can be enhanced at intermediate frequencies for near-field interacting particles. We observes that this enhancement can be very large compared to what can be achieved in static limit. This would appear to indicate that the maximum variation of $T_2(t)$ in two-body systems may be amplified or attenuated by adding another nanoparticle in between and can be engineered by geometrical arrangement. To see if the presence of the third particle can delay or accelerate the temporal response of particle 2, the frequency dependent phase lag difference caused by the presence of the third particle, i.e., $\Delta\varphi=\varphi_{21}^{(1)}|_{\textrm{3-body}}-\varphi_{21}^{(1)}|_{\textrm{2-body}}$, is presented in Fig.~(\ref{Figure.6}b) for various separation distances. 
 We observe that the response is immediate in the static limit, i. e., $\Delta\varphi=\varphi_{21}^{(1)}|_{\textrm{3-body}}=\varphi_{21}^{(1)}|_{\textrm{2-body}}=0$. However, the response of particle 2, could be delayed or accelerated at higher frequency input powers due to the presence of third particle.  It is clear that for sufficiently small separation distances the phase lag difference $\Delta\varphi$ is negative which means that the presence of particle 3 accelerated the response of $T_2(t)$ compared to the two-body case. On the other side, when the separation distances are large we may have $\Delta\varphi>0$, which implies that the temporal evolution of $T_2(t)$  is delayed compared to the two-body case.  
\section{Conclusion}\label{sec4}
In this work we have introduced a theoretical
framework to investigate the dynamics of radiative heat transfer in many-body systems. 
The complex response matrices were quantified and successfully applied to estimate the temperature dynamics in particle systems. The study of response function in a three-body system suggests that the dynamic response of many body systems to an external input power may be amplified, attenuated, delayed or accelerated due to geometric arrangement.  Moreover, our calculation shows that  the use of  time-dependent external energy  sources can be actively exploited to produce a desired time dependent temperature response in a collection of nanoparticles which can be engineered by their geometrical distribution. In addition, eigenvalues and eigenvectors of the response matrix can provide insights into the direction of fast and slow heat transfer at the nanoscale. It should be emphasized that if the input powers are very large, higher order response terms become important and the dynamic of temperatures cannot adequately be described just by its linear response function. However, the calculated linear response to harmonic input powers allows the study of temperature dynamics in the presence of   weak aperiodic input powers.
\begin{acknowledgements}
The author would like to thank P. B. Abdallah for helpful discussions.
\end{acknowledgements}
\section*{reference}

\begin{thebibliography}{58}%
\makeatletter
\providecommand \@ifxundefined [1]{%
 \@ifx{#1\undefined}
}%
\providecommand \@ifnum [1]{%
 \ifnum #1\expandafter \@firstoftwo
 \else \expandafter \@secondoftwo
 \fi
}%
\providecommand \@ifx [1]{%
 \ifx #1\expandafter \@firstoftwo
 \else \expandafter \@secondoftwo
 \fi
}%
\providecommand \natexlab [1]{#1}%
\providecommand \enquote  [1]{``#1''}%
\providecommand \bibnamefont  [1]{#1}%
\providecommand \bibfnamefont [1]{#1}%
\providecommand \citenamefont [1]{#1}%
\providecommand \href@noop [0]{\@secondoftwo}%
\providecommand \href [0]{\begingroup \@sanitize@url \@href}%
\providecommand \@href[1]{\@@startlink{#1}\@@href}%
\providecommand \@@href[1]{\endgroup#1\@@endlink}%
\providecommand \@sanitize@url [0]{\catcode `\\12\catcode `\$12\catcode
  `\&12\catcode `\#12\catcode `\^12\catcode `\_12\catcode `\%12\relax}%
\providecommand \@@startlink[1]{}%
\providecommand \@@endlink[0]{}%
\providecommand \url  [0]{\begingroup\@sanitize@url \@url }%
\providecommand \@url [1]{\endgroup\@href {#1}{\urlprefix }}%
\providecommand \urlprefix  [0]{URL }%
\providecommand \Eprint [0]{\href }%
\providecommand \doibase [0]{http://dx.doi.org/}%
\providecommand \selectlanguage [0]{\@gobble}%
\providecommand \bibinfo  [0]{\@secondoftwo}%
\providecommand \bibfield  [0]{\@secondoftwo}%
\providecommand \translation [1]{[#1]}%
\providecommand \BibitemOpen [0]{}%
\providecommand \bibitemStop [0]{}%
\providecommand \bibitemNoStop [0]{.\EOS\space}%
\providecommand \EOS [0]{\spacefactor3000\relax}%
\providecommand \BibitemShut  [1]{\csname bibitem#1\endcsname}%
\let\auto@bib@innerbib\@empty
\bibitem [{\citenamefont {Cravalho}\ \emph {et~al.}(1967)\citenamefont
  {Cravalho}, \citenamefont {Tien},\ and\ \citenamefont {Caren}}]{Cravalho}%
  \BibitemOpen
  \bibfield  {author} {\bibinfo {author} {\bibfnamefont {E.~G.}\ \bibnamefont
  {Cravalho}}, \bibinfo {author} {\bibfnamefont {C.~L.}\ \bibnamefont {Tien}},
  \ and\ \bibinfo {author} {\bibfnamefont {R.~P.}\ \bibnamefont {Caren}},\
  }\href {\doibase 10.1115/1.3614396} {\bibfield  {journal} {\bibinfo
  {journal} {Journal of Heat Transfer}\ }\textbf {\bibinfo {volume} {89}},\
  \bibinfo {pages} {351} (\bibinfo {year} {1967})},\ \Eprint
  {http://arxiv.org/abs/https://asmedigitalcollection.asme.org/heattransfer/article-pdf/89/4/351/5698484/351\_1.pdf}
  {https://asmedigitalcollection.asme.org/heattransfer/article-pdf/89/4/351/5698484/351\_1.pdf}
  \BibitemShut {NoStop}%
\bibitem [{\citenamefont {Boehm}\ and\ \citenamefont {Tien}(1970)}]{Boehm}%
  \BibitemOpen
  \bibfield  {author} {\bibinfo {author} {\bibfnamefont {R.~F.}\ \bibnamefont
  {Boehm}}\ and\ \bibinfo {author} {\bibfnamefont {C.~L.}\ \bibnamefont
  {Tien}},\ }\href {\doibase 10.1115/1.3449676} {\bibfield  {journal} {\bibinfo
   {journal} {Journal of Heat Transfer}\ }\textbf {\bibinfo {volume} {92}},\
  \bibinfo {pages} {405} (\bibinfo {year} {1970})},\ \Eprint
  {http://arxiv.org/abs/https://asmedigitalcollection.asme.org/heattransfer/article-pdf/92/3/405/5577224/405\_1.pdf}
  {https://asmedigitalcollection.asme.org/heattransfer/article-pdf/92/3/405/5577224/405\_1.pdf}
  \BibitemShut {NoStop}%
\bibitem [{\citenamefont {Polder}\ and\ \citenamefont
  {Van~Hove}(1971)}]{PhysRevB.4.3303}%
  \BibitemOpen
  \bibfield  {author} {\bibinfo {author} {\bibfnamefont {D.}~\bibnamefont
  {Polder}}\ and\ \bibinfo {author} {\bibfnamefont {M.}~\bibnamefont
  {Van~Hove}},\ }\href {\doibase 10.1103/PhysRevB.4.3303} {\bibfield  {journal}
  {\bibinfo  {journal} {Phys. Rev. B}\ }\textbf {\bibinfo {volume} {4}},\
  \bibinfo {pages} {3303} (\bibinfo {year} {1971})}\BibitemShut {NoStop}%
\bibitem [{\citenamefont {Rytov}(1959)}]{rytov1959theory}%
  \BibitemOpen
  \bibfield  {author} {\bibinfo {author} {\bibfnamefont {S.}~\bibnamefont
  {Rytov}},\ }\href {https://books.google.com/books?id=Peh4SwAACAAJ} {\emph
  {\bibinfo {title} {Theory of Electric Fluctuations and Thermal Radiation}}},\
  AFCRC-TR\ (\bibinfo  {publisher} {Air Force Cambridge Research Center},\
  \bibinfo {year} {1959})\BibitemShut {NoStop}%
\bibitem [{\citenamefont {Ott}\ and\ \citenamefont
  {Biehs}(2020{\natexlab{a}})}]{PhysRevB.102.115417}%
  \BibitemOpen
  \bibfield  {author} {\bibinfo {author} {\bibfnamefont {A.}~\bibnamefont
  {Ott}}\ and\ \bibinfo {author} {\bibfnamefont {S.-A.}\ \bibnamefont
  {Biehs}},\ }\href {\doibase 10.1103/PhysRevB.102.115417} {\bibfield
  {journal} {\bibinfo  {journal} {Phys. Rev. B}\ }\textbf {\bibinfo {volume}
  {102}},\ \bibinfo {pages} {115417} (\bibinfo {year}
  {2020}{\natexlab{a}})}\BibitemShut {NoStop}%
\bibitem [{\citenamefont {Moncada-Villa}\ and\ \citenamefont
  {Cuevas}(2020{\natexlab{a}})}]{moncadavilla2020normal}%
  \BibitemOpen
  \bibfield  {author} {\bibinfo {author} {\bibfnamefont {E.}~\bibnamefont
  {Moncada-Villa}}\ and\ \bibinfo {author} {\bibfnamefont {J.~C.}\ \bibnamefont
  {Cuevas}},\ }\href@noop {} {\enquote {\bibinfo {title} {Normal
  metal-superconductor near-field thermal diodes and transistors},}\ }
  (\bibinfo {year} {2020}{\natexlab{a}}),\ \Eprint
  {http://arxiv.org/abs/2011.10026} {arXiv:2011.10026 [cond-mat.supr-con]}
  \BibitemShut {NoStop}%
\bibitem [{\citenamefont {Luo}\ \emph {et~al.}(2019)\citenamefont {Luo},
  \citenamefont {Dong}, \citenamefont {Zhao}, \citenamefont {Liu},\ and\
  \citenamefont {Antezza}}]{PhysRevB.99.134207}%
  \BibitemOpen
  \bibfield  {author} {\bibinfo {author} {\bibfnamefont {M.}~\bibnamefont
  {Luo}}, \bibinfo {author} {\bibfnamefont {J.}~\bibnamefont {Dong}}, \bibinfo
  {author} {\bibfnamefont {J.}~\bibnamefont {Zhao}}, \bibinfo {author}
  {\bibfnamefont {L.}~\bibnamefont {Liu}}, \ and\ \bibinfo {author}
  {\bibfnamefont {M.}~\bibnamefont {Antezza}},\ }\href {\doibase
  10.1103/PhysRevB.99.134207} {\bibfield  {journal} {\bibinfo  {journal} {Phys.
  Rev. B}\ }\textbf {\bibinfo {volume} {99}},\ \bibinfo {pages} {134207}
  (\bibinfo {year} {2019})}\BibitemShut {NoStop}%
\bibitem [{\citenamefont {Ramezan~Choubdar}\ and\ \citenamefont
  {Nikbakht}(2016)}]{Choubdar}%
  \BibitemOpen
  \bibfield  {author} {\bibinfo {author} {\bibfnamefont {O.}~\bibnamefont
  {Ramezan~Choubdar}}\ and\ \bibinfo {author} {\bibfnamefont {M.}~\bibnamefont
  {Nikbakht}},\ }\href {\doibase 10.1063/1.4964698} {\bibfield  {journal}
  {\bibinfo  {journal} {Journal of Applied Physics}\ }\textbf {\bibinfo
  {volume} {120}},\ \bibinfo {pages} {144303} (\bibinfo {year} {2016})},\
  \Eprint {http://arxiv.org/abs/https://doi.org/10.1063/1.4964698}
  {https://doi.org/10.1063/1.4964698} \BibitemShut {NoStop}%
\bibitem [{\citenamefont {Nikbakht}(2018)}]{NIKBAKHT2018164}%
  \BibitemOpen
  \bibfield  {author} {\bibinfo {author} {\bibfnamefont {M.}~\bibnamefont
  {Nikbakht}},\ }\href {\doibase https://doi.org/10.1016/j.jqsrt.2018.10.005}
  {\bibfield  {journal} {\bibinfo  {journal} {Journal of Quantitative
  Spectroscopy and Radiative Transfer}\ }\textbf {\bibinfo {volume} {221}},\
  \bibinfo {pages} {164} (\bibinfo {year} {2018})}\BibitemShut {NoStop}%
\bibitem [{\citenamefont {Luo}\ \emph {et~al.}(2020)\citenamefont {Luo},
  \citenamefont {Zhao},\ and\ \citenamefont {Antezza}}]{doi:10.1063/5.0018329}%
  \BibitemOpen
  \bibfield  {author} {\bibinfo {author} {\bibfnamefont {M.}~\bibnamefont
  {Luo}}, \bibinfo {author} {\bibfnamefont {J.}~\bibnamefont {Zhao}}, \ and\
  \bibinfo {author} {\bibfnamefont {M.}~\bibnamefont {Antezza}},\ }\href
  {\doibase 10.1063/5.0018329} {\bibfield  {journal} {\bibinfo  {journal}
  {Applied Physics Letters}\ }\textbf {\bibinfo {volume} {117}},\ \bibinfo
  {pages} {053901} (\bibinfo {year} {2020})},\ \Eprint
  {http://arxiv.org/abs/https://doi.org/10.1063/5.0018329}
  {https://doi.org/10.1063/5.0018329} \BibitemShut {NoStop}%
\bibitem [{\citenamefont {Ben-Abdallah}\ \emph {et~al.}(2011)\citenamefont
  {Ben-Abdallah}, \citenamefont {Biehs},\ and\ \citenamefont
  {Joulain}}]{PhysRevLett.107.114301}%
  \BibitemOpen
  \bibfield  {author} {\bibinfo {author} {\bibfnamefont {P.}~\bibnamefont
  {Ben-Abdallah}}, \bibinfo {author} {\bibfnamefont {S.-A.}\ \bibnamefont
  {Biehs}}, \ and\ \bibinfo {author} {\bibfnamefont {K.}~\bibnamefont
  {Joulain}},\ }\href {\doibase 10.1103/PhysRevLett.107.114301} {\bibfield
  {journal} {\bibinfo  {journal} {Phys. Rev. Lett.}\ }\textbf {\bibinfo
  {volume} {107}},\ \bibinfo {pages} {114301} (\bibinfo {year}
  {2011})}\BibitemShut {NoStop}%
\bibitem [{\citenamefont {Latella}\ \emph {et~al.}(2017)\citenamefont
  {Latella}, \citenamefont {Ben-Abdallah}, \citenamefont {Biehs}, \citenamefont
  {Antezza},\ and\ \citenamefont {Messina}}]{PhysRevB.95.205404}%
  \BibitemOpen
  \bibfield  {author} {\bibinfo {author} {\bibfnamefont {I.}~\bibnamefont
  {Latella}}, \bibinfo {author} {\bibfnamefont {P.}~\bibnamefont
  {Ben-Abdallah}}, \bibinfo {author} {\bibfnamefont {S.-A.}\ \bibnamefont
  {Biehs}}, \bibinfo {author} {\bibfnamefont {M.}~\bibnamefont {Antezza}}, \
  and\ \bibinfo {author} {\bibfnamefont {R.}~\bibnamefont {Messina}},\ }\href
  {\doibase 10.1103/PhysRevB.95.205404} {\bibfield  {journal} {\bibinfo
  {journal} {Phys. Rev. B}\ }\textbf {\bibinfo {volume} {95}},\ \bibinfo
  {pages} {205404} (\bibinfo {year} {2017})}\BibitemShut {NoStop}%
\bibitem [{\citenamefont {Nikbakht}(2014)}]{manybody}%
  \BibitemOpen
  \bibfield  {author} {\bibinfo {author} {\bibfnamefont {M.}~\bibnamefont
  {Nikbakht}},\ }\href {\doibase 10.1063/1.4894622} {\bibfield  {journal}
  {\bibinfo  {journal} {Journal of Applied Physics}\ }\textbf {\bibinfo
  {volume} {116}},\ \bibinfo {pages} {094307} (\bibinfo {year} {2014})},\
  \Eprint {http://arxiv.org/abs/https://doi.org/10.1063/1.4894622}
  {https://doi.org/10.1063/1.4894622} \BibitemShut {NoStop}%
\bibitem [{\citenamefont {Nikbakht}(2017)}]{PhysRevB.96.125436}%
  \BibitemOpen
  \bibfield  {author} {\bibinfo {author} {\bibfnamefont {M.}~\bibnamefont
  {Nikbakht}},\ }\href {\doibase 10.1103/PhysRevB.96.125436} {\bibfield
  {journal} {\bibinfo  {journal} {Phys. Rev. B}\ }\textbf {\bibinfo {volume}
  {96}},\ \bibinfo {pages} {125436} (\bibinfo {year} {2017})}\BibitemShut
  {NoStop}%
\bibitem [{\citenamefont {Abraham~Ekeroth}\ \emph {et~al.}(2017)\citenamefont
  {Abraham~Ekeroth}, \citenamefont {Garc\'{\i}a-Mart\'{\i}n},\ and\
  \citenamefont {Cuevas}}]{PhysRevB.95.235428}%
  \BibitemOpen
  \bibfield  {author} {\bibinfo {author} {\bibfnamefont {R.~M.}\ \bibnamefont
  {Abraham~Ekeroth}}, \bibinfo {author} {\bibfnamefont {A.}~\bibnamefont
  {Garc\'{\i}a-Mart\'{\i}n}}, \ and\ \bibinfo {author} {\bibfnamefont {J.~C.}\
  \bibnamefont {Cuevas}},\ }\href {\doibase 10.1103/PhysRevB.95.235428}
  {\bibfield  {journal} {\bibinfo  {journal} {Phys. Rev. B}\ }\textbf {\bibinfo
  {volume} {95}},\ \bibinfo {pages} {235428} (\bibinfo {year}
  {2017})}\BibitemShut {NoStop}%
\bibitem [{\citenamefont {Dong}\ \emph {et~al.}(2017)\citenamefont {Dong},
  \citenamefont {Zhao},\ and\ \citenamefont {Liu}}]{PhysRevB.95.125411}%
  \BibitemOpen
  \bibfield  {author} {\bibinfo {author} {\bibfnamefont {J.}~\bibnamefont
  {Dong}}, \bibinfo {author} {\bibfnamefont {J.}~\bibnamefont {Zhao}}, \ and\
  \bibinfo {author} {\bibfnamefont {L.}~\bibnamefont {Liu}},\ }\href {\doibase
  10.1103/PhysRevB.95.125411} {\bibfield  {journal} {\bibinfo  {journal} {Phys.
  Rev. B}\ }\textbf {\bibinfo {volume} {95}},\ \bibinfo {pages} {125411}
  (\bibinfo {year} {2017})}\BibitemShut {NoStop}%
\bibitem [{\citenamefont {Song}\ \emph {et~al.}(2021)\citenamefont {Song},
  \citenamefont {Cheng}, \citenamefont {Zhang}, \citenamefont {Lu},
  \citenamefont {Zhou}, \citenamefont {Luo},\ and\ \citenamefont
  {Hu}}]{10.1088/1361-6633/abe52b}%
  \BibitemOpen
  \bibfield  {author} {\bibinfo {author} {\bibfnamefont {J.}~\bibnamefont
  {Song}}, \bibinfo {author} {\bibfnamefont {Q.}~\bibnamefont {Cheng}},
  \bibinfo {author} {\bibfnamefont {B.}~\bibnamefont {Zhang}}, \bibinfo
  {author} {\bibfnamefont {L.}~\bibnamefont {Lu}}, \bibinfo {author}
  {\bibfnamefont {X.}~\bibnamefont {Zhou}}, \bibinfo {author} {\bibfnamefont
  {Z.}~\bibnamefont {Luo}}, \ and\ \bibinfo {author} {\bibfnamefont
  {R.}~\bibnamefont {Hu}},\ }\href
  {http://iopscience.iop.org/article/10.1088/1361-6633/abe52b} {\bibfield
  {journal} {\bibinfo  {journal} {Reports on Progress in Physics}\ } (\bibinfo
  {year} {2021})}\BibitemShut {NoStop}%
\bibitem [{\citenamefont {Tervo}\ \emph {et~al.}(2019)\citenamefont {Tervo},
  \citenamefont {Francoeur}, \citenamefont {Cola},\ and\ \citenamefont
  {Zhang}}]{PhysRevB.100.205422}%
  \BibitemOpen
  \bibfield  {author} {\bibinfo {author} {\bibfnamefont {E.}~\bibnamefont
  {Tervo}}, \bibinfo {author} {\bibfnamefont {M.}~\bibnamefont {Francoeur}},
  \bibinfo {author} {\bibfnamefont {B.}~\bibnamefont {Cola}}, \ and\ \bibinfo
  {author} {\bibfnamefont {Z.}~\bibnamefont {Zhang}},\ }\href {\doibase
  10.1103/PhysRevB.100.205422} {\bibfield  {journal} {\bibinfo  {journal}
  {Phys. Rev. B}\ }\textbf {\bibinfo {volume} {100}},\ \bibinfo {pages}
  {205422} (\bibinfo {year} {2019})}\BibitemShut {NoStop}%
\bibitem [{\citenamefont {Asheichyk}\ \emph {et~al.}(2017)\citenamefont
  {Asheichyk}, \citenamefont {M\"uller},\ and\ \citenamefont
  {Kr\"uger}}]{PhysRevB.96.155402}%
  \BibitemOpen
  \bibfield  {author} {\bibinfo {author} {\bibfnamefont {K.}~\bibnamefont
  {Asheichyk}}, \bibinfo {author} {\bibfnamefont {B.}~\bibnamefont {M\"uller}},
  \ and\ \bibinfo {author} {\bibfnamefont {M.}~\bibnamefont {Kr\"uger}},\
  }\href {\doibase 10.1103/PhysRevB.96.155402} {\bibfield  {journal} {\bibinfo
  {journal} {Phys. Rev. B}\ }\textbf {\bibinfo {volume} {96}},\ \bibinfo
  {pages} {155402} (\bibinfo {year} {2017})}\BibitemShut {NoStop}%
\bibitem [{\citenamefont {Ben-Abdallah}\ and\ \citenamefont
  {Biehs}(2013)}]{doi:10.1063/1.4829618}%
  \BibitemOpen
  \bibfield  {author} {\bibinfo {author} {\bibfnamefont {P.}~\bibnamefont
  {Ben-Abdallah}}\ and\ \bibinfo {author} {\bibfnamefont {S.-A.}\ \bibnamefont
  {Biehs}},\ }\href {\doibase 10.1063/1.4829618} {\bibfield  {journal}
  {\bibinfo  {journal} {Applied Physics Letters}\ }\textbf {\bibinfo {volume}
  {103}},\ \bibinfo {pages} {191907} (\bibinfo {year} {2013})},\ \Eprint
  {http://arxiv.org/abs/https://doi.org/10.1063/1.4829618}
  {https://doi.org/10.1063/1.4829618} \BibitemShut {NoStop}%
\bibitem [{\citenamefont {Ben-Abdallah}\ and\ \citenamefont
  {Biehs}(2015)}]{doi:10.1063/1.4915138}%
  \BibitemOpen
  \bibfield  {author} {\bibinfo {author} {\bibfnamefont {P.}~\bibnamefont
  {Ben-Abdallah}}\ and\ \bibinfo {author} {\bibfnamefont {S.-A.}\ \bibnamefont
  {Biehs}},\ }\href {\doibase 10.1063/1.4915138} {\bibfield  {journal}
  {\bibinfo  {journal} {AIP Advances}\ }\textbf {\bibinfo {volume} {5}},\
  \bibinfo {pages} {053502} (\bibinfo {year} {2015})},\ \Eprint
  {http://arxiv.org/abs/https://doi.org/10.1063/1.4915138}
  {https://doi.org/10.1063/1.4915138} \BibitemShut {NoStop}%
\bibitem [{\citenamefont {Fiorino}\ \emph {et~al.}(2018)\citenamefont
  {Fiorino}, \citenamefont {Thompson}, \citenamefont {Zhu}, \citenamefont
  {Mittapally}, \citenamefont {Biehs}, \citenamefont {Bezencenet},
  \citenamefont {El-Bondry}, \citenamefont {Bansropun}, \citenamefont
  {Ben-Abdallah}, \citenamefont {Meyhofer},\ and\ \citenamefont
  {Reddy}}]{doi:10.1021/acsnano.8b01645}%
  \BibitemOpen
  \bibfield  {author} {\bibinfo {author} {\bibfnamefont {A.}~\bibnamefont
  {Fiorino}}, \bibinfo {author} {\bibfnamefont {D.}~\bibnamefont {Thompson}},
  \bibinfo {author} {\bibfnamefont {L.}~\bibnamefont {Zhu}}, \bibinfo {author}
  {\bibfnamefont {R.}~\bibnamefont {Mittapally}}, \bibinfo {author}
  {\bibfnamefont {S.-A.}\ \bibnamefont {Biehs}}, \bibinfo {author}
  {\bibfnamefont {O.}~\bibnamefont {Bezencenet}}, \bibinfo {author}
  {\bibfnamefont {N.}~\bibnamefont {El-Bondry}}, \bibinfo {author}
  {\bibfnamefont {S.}~\bibnamefont {Bansropun}}, \bibinfo {author}
  {\bibfnamefont {P.}~\bibnamefont {Ben-Abdallah}}, \bibinfo {author}
  {\bibfnamefont {E.}~\bibnamefont {Meyhofer}}, \ and\ \bibinfo {author}
  {\bibfnamefont {P.}~\bibnamefont {Reddy}},\ }\href {\doibase
  10.1021/acsnano.8b01645} {\bibfield  {journal} {\bibinfo  {journal} {ACS
  Nano}\ }\textbf {\bibinfo {volume} {12}},\ \bibinfo {pages} {5774} (\bibinfo
  {year} {2018})},\ \bibinfo {note} {pMID: 29790344},\ \Eprint
  {http://arxiv.org/abs/https://doi.org/10.1021/acsnano.8b01645}
  {https://doi.org/10.1021/acsnano.8b01645} \BibitemShut {NoStop}%
\bibitem [{\citenamefont {Ben-Abdallah}(2019)}]{PhysRevB.99.201406}%
  \BibitemOpen
  \bibfield  {author} {\bibinfo {author} {\bibfnamefont {P.}~\bibnamefont
  {Ben-Abdallah}},\ }\href {\doibase 10.1103/PhysRevB.99.201406} {\bibfield
  {journal} {\bibinfo  {journal} {Phys. Rev. B}\ }\textbf {\bibinfo {volume}
  {99}},\ \bibinfo {pages} {201406} (\bibinfo {year} {2019})}\BibitemShut
  {NoStop}%
\bibitem [{\citenamefont {Ott}\ and\ \citenamefont
  {Biehs}(2020{\natexlab{b}})}]{PhysRevB.101.155428}%
  \BibitemOpen
  \bibfield  {author} {\bibinfo {author} {\bibfnamefont {A.}~\bibnamefont
  {Ott}}\ and\ \bibinfo {author} {\bibfnamefont {S.-A.}\ \bibnamefont
  {Biehs}},\ }\href {\doibase 10.1103/PhysRevB.101.155428} {\bibfield
  {journal} {\bibinfo  {journal} {Phys. Rev. B}\ }\textbf {\bibinfo {volume}
  {101}},\ \bibinfo {pages} {155428} (\bibinfo {year}
  {2020}{\natexlab{b}})}\BibitemShut {NoStop}%
\bibitem [{\citenamefont {Zhu}\ and\ \citenamefont
  {Fan}(2016)}]{PhysRevLett.117.134303}%
  \BibitemOpen
  \bibfield  {author} {\bibinfo {author} {\bibfnamefont {L.}~\bibnamefont
  {Zhu}}\ and\ \bibinfo {author} {\bibfnamefont {S.}~\bibnamefont {Fan}},\
  }\href {\doibase 10.1103/PhysRevLett.117.134303} {\bibfield  {journal}
  {\bibinfo  {journal} {Phys. Rev. Lett.}\ }\textbf {\bibinfo {volume} {117}},\
  \bibinfo {pages} {134303} (\bibinfo {year} {2016})}\BibitemShut {NoStop}%
\bibitem [{\citenamefont {Zhu}\ \emph {et~al.}(2018)\citenamefont {Zhu},
  \citenamefont {Guo},\ and\ \citenamefont {Fan}}]{PhysRevB.97.094302}%
  \BibitemOpen
  \bibfield  {author} {\bibinfo {author} {\bibfnamefont {L.}~\bibnamefont
  {Zhu}}, \bibinfo {author} {\bibfnamefont {Y.}~\bibnamefont {Guo}}, \ and\
  \bibinfo {author} {\bibfnamefont {S.}~\bibnamefont {Fan}},\ }\href {\doibase
  10.1103/PhysRevB.97.094302} {\bibfield  {journal} {\bibinfo  {journal} {Phys.
  Rev. B}\ }\textbf {\bibinfo {volume} {97}},\ \bibinfo {pages} {094302}
  (\bibinfo {year} {2018})}\BibitemShut {NoStop}%
\bibitem [{\citenamefont {Guo}\ and\ \citenamefont
  {Fan}(2020)}]{PhysRevB.102.085401}%
  \BibitemOpen
  \bibfield  {author} {\bibinfo {author} {\bibfnamefont {C.}~\bibnamefont
  {Guo}}\ and\ \bibinfo {author} {\bibfnamefont {S.}~\bibnamefont {Fan}},\
  }\href {\doibase 10.1103/PhysRevB.102.085401} {\bibfield  {journal} {\bibinfo
   {journal} {Phys. Rev. B}\ }\textbf {\bibinfo {volume} {102}},\ \bibinfo
  {pages} {085401} (\bibinfo {year} {2020})}\BibitemShut {NoStop}%
\bibitem [{\citenamefont {Zhang}\ \emph
  {et~al.}(2020{\natexlab{a}})\citenamefont {Zhang}, \citenamefont {Zhou},
  \citenamefont {Yi},\ and\ \citenamefont {Tan}}]{PhysRevApplied.13.034021}%
  \BibitemOpen
  \bibfield  {author} {\bibinfo {author} {\bibfnamefont {Y.}~\bibnamefont
  {Zhang}}, \bibinfo {author} {\bibfnamefont {C.-L.}\ \bibnamefont {Zhou}},
  \bibinfo {author} {\bibfnamefont {H.-L.}\ \bibnamefont {Yi}}, \ and\ \bibinfo
  {author} {\bibfnamefont {H.-P.}\ \bibnamefont {Tan}},\ }\href {\doibase
  10.1103/PhysRevApplied.13.034021} {\bibfield  {journal} {\bibinfo  {journal}
  {Phys. Rev. Applied}\ }\textbf {\bibinfo {volume} {13}},\ \bibinfo {pages}
  {034021} (\bibinfo {year} {2020}{\natexlab{a}})}\BibitemShut {NoStop}%
\bibitem [{\citenamefont {Messina}\ \emph {et~al.}(2013)\citenamefont
  {Messina}, \citenamefont {Tschikin}, \citenamefont {Biehs},\ and\
  \citenamefont {Ben-Abdallah}}]{PhysRevB.88.104307}%
  \BibitemOpen
  \bibfield  {author} {\bibinfo {author} {\bibfnamefont {R.}~\bibnamefont
  {Messina}}, \bibinfo {author} {\bibfnamefont {M.}~\bibnamefont {Tschikin}},
  \bibinfo {author} {\bibfnamefont {S.-A.}\ \bibnamefont {Biehs}}, \ and\
  \bibinfo {author} {\bibfnamefont {P.}~\bibnamefont {Ben-Abdallah}},\ }\href
  {\doibase 10.1103/PhysRevB.88.104307} {\bibfield  {journal} {\bibinfo
  {journal} {Phys. Rev. B}\ }\textbf {\bibinfo {volume} {88}},\ \bibinfo
  {pages} {104307} (\bibinfo {year} {2013})}\BibitemShut {NoStop}%
\bibitem [{\citenamefont {Nikbakht}(2015)}]{Nikbakht_2015}%
  \BibitemOpen
  \bibfield  {author} {\bibinfo {author} {\bibfnamefont {M.}~\bibnamefont
  {Nikbakht}},\ }\href {\doibase 10.1209/0295-5075/110/14004} {\bibfield
  {journal} {\bibinfo  {journal} {{EPL} (Europhysics Letters)}\ }\textbf
  {\bibinfo {volume} {110}},\ \bibinfo {pages} {14004} (\bibinfo {year}
  {2015})}\BibitemShut {NoStop}%
\bibitem [{\citenamefont {Kubytskyi}\ \emph {et~al.}(2014)\citenamefont
  {Kubytskyi}, \citenamefont {Biehs},\ and\ \citenamefont
  {Ben-Abdallah}}]{PhysRevLett.113.074301}%
  \BibitemOpen
  \bibfield  {author} {\bibinfo {author} {\bibfnamefont {V.}~\bibnamefont
  {Kubytskyi}}, \bibinfo {author} {\bibfnamefont {S.-A.}\ \bibnamefont
  {Biehs}}, \ and\ \bibinfo {author} {\bibfnamefont {P.}~\bibnamefont
  {Ben-Abdallah}},\ }\href {\doibase 10.1103/PhysRevLett.113.074301} {\bibfield
   {journal} {\bibinfo  {journal} {Phys. Rev. Lett.}\ }\textbf {\bibinfo
  {volume} {113}},\ \bibinfo {pages} {074301} (\bibinfo {year}
  {2014})}\BibitemShut {NoStop}%
\bibitem [{\citenamefont {Zolghadr}\ and\ \citenamefont
  {Nikbakht}(2020)}]{PhysRevB.102.035433}%
  \BibitemOpen
  \bibfield  {author} {\bibinfo {author} {\bibfnamefont {N.}~\bibnamefont
  {Zolghadr}}\ and\ \bibinfo {author} {\bibfnamefont {M.}~\bibnamefont
  {Nikbakht}},\ }\href {\doibase 10.1103/PhysRevB.102.035433} {\bibfield
  {journal} {\bibinfo  {journal} {Phys. Rev. B}\ }\textbf {\bibinfo {volume}
  {102}},\ \bibinfo {pages} {035433} (\bibinfo {year} {2020})}\BibitemShut
  {NoStop}%
\bibitem [{\citenamefont {Qu}\ \emph {et~al.}(2021)\citenamefont {Qu},
  \citenamefont {Zhang}, \citenamefont {Fang},\ and\ \citenamefont
  {Yi}}]{QU2021107404}%
  \BibitemOpen
  \bibfield  {author} {\bibinfo {author} {\bibfnamefont {L.}~\bibnamefont
  {Qu}}, \bibinfo {author} {\bibfnamefont {Y.}~\bibnamefont {Zhang}}, \bibinfo
  {author} {\bibfnamefont {J.-L.}\ \bibnamefont {Fang}}, \ and\ \bibinfo
  {author} {\bibfnamefont {H.-L.}\ \bibnamefont {Yi}},\ }\href {\doibase
  https://doi.org/10.1016/j.jqsrt.2020.107404} {\bibfield  {journal} {\bibinfo
  {journal} {Journal of Quantitative Spectroscopy and Radiative Transfer}\
  }\textbf {\bibinfo {volume} {258}},\ \bibinfo {pages} {107404} (\bibinfo
  {year} {2021})}\BibitemShut {NoStop}%
\bibitem [{\citenamefont {Moncada-Villa}\ and\ \citenamefont
  {Cuevas}(2020{\natexlab{b}})}]{PhysRevB.101.085411}%
  \BibitemOpen
  \bibfield  {author} {\bibinfo {author} {\bibfnamefont {E.}~\bibnamefont
  {Moncada-Villa}}\ and\ \bibinfo {author} {\bibfnamefont {J.~C.}\ \bibnamefont
  {Cuevas}},\ }\href {\doibase 10.1103/PhysRevB.101.085411} {\bibfield
  {journal} {\bibinfo  {journal} {Phys. Rev. B}\ }\textbf {\bibinfo {volume}
  {101}},\ \bibinfo {pages} {085411} (\bibinfo {year}
  {2020}{\natexlab{b}})}\BibitemShut {NoStop}%
\bibitem [{\citenamefont {Zhang}\ \emph
  {et~al.}(2020{\natexlab{b}})\citenamefont {Zhang}, \citenamefont {Zhou},
  \citenamefont {Qu},\ and\ \citenamefont {Yi}}]{doi:10.1063/1.5145224}%
  \BibitemOpen
  \bibfield  {author} {\bibinfo {author} {\bibfnamefont {Y.}~\bibnamefont
  {Zhang}}, \bibinfo {author} {\bibfnamefont {C.-L.}\ \bibnamefont {Zhou}},
  \bibinfo {author} {\bibfnamefont {L.}~\bibnamefont {Qu}}, \ and\ \bibinfo
  {author} {\bibfnamefont {H.-L.}\ \bibnamefont {Yi}},\ }\href {\doibase
  10.1063/1.5145224} {\bibfield  {journal} {\bibinfo  {journal} {Applied
  Physics Letters}\ }\textbf {\bibinfo {volume} {116}},\ \bibinfo {pages}
  {151101} (\bibinfo {year} {2020}{\natexlab{b}})},\ \Eprint
  {http://arxiv.org/abs/https://doi.org/10.1063/1.5145224}
  {https://doi.org/10.1063/1.5145224} \BibitemShut {NoStop}%
\bibitem [{\citenamefont {Zundel}\ and\ \citenamefont
  {Manjavacas}(2020)}]{PhysRevApplied.13.054054}%
  \BibitemOpen
  \bibfield  {author} {\bibinfo {author} {\bibfnamefont {L.}~\bibnamefont
  {Zundel}}\ and\ \bibinfo {author} {\bibfnamefont {A.}~\bibnamefont
  {Manjavacas}},\ }\href {\doibase 10.1103/PhysRevApplied.13.054054} {\bibfield
   {journal} {\bibinfo  {journal} {Phys. Rev. Applied}\ }\textbf {\bibinfo
  {volume} {13}},\ \bibinfo {pages} {054054} (\bibinfo {year}
  {2020})}\BibitemShut {NoStop}%
\bibitem [{\citenamefont {Dyakov}\ \emph {et~al.}(2015)\citenamefont {Dyakov},
  \citenamefont {Dai}, \citenamefont {Yan},\ and\ \citenamefont
  {Qiu}}]{Dyakov_2015}%
  \BibitemOpen
  \bibfield  {author} {\bibinfo {author} {\bibfnamefont {S.~A.}\ \bibnamefont
  {Dyakov}}, \bibinfo {author} {\bibfnamefont {J.}~\bibnamefont {Dai}},
  \bibinfo {author} {\bibfnamefont {M.}~\bibnamefont {Yan}}, \ and\ \bibinfo
  {author} {\bibfnamefont {M.}~\bibnamefont {Qiu}},\ }\href {\doibase
  10.1088/0022-3727/48/30/305104} {\bibfield  {journal} {\bibinfo  {journal}
  {Journal of Physics D: Applied Physics}\ }\textbf {\bibinfo {volume} {48}},\
  \bibinfo {pages} {305104} (\bibinfo {year} {2015})}\BibitemShut {NoStop}%
\bibitem [{\citenamefont {Ben-Abdallah}\ and\ \citenamefont
  {Biehs}(2014)}]{PhysRevLett.112.044301}%
  \BibitemOpen
  \bibfield  {author} {\bibinfo {author} {\bibfnamefont {P.}~\bibnamefont
  {Ben-Abdallah}}\ and\ \bibinfo {author} {\bibfnamefont {S.-A.}\ \bibnamefont
  {Biehs}},\ }\href {\doibase 10.1103/PhysRevLett.112.044301} {\bibfield
  {journal} {\bibinfo  {journal} {Phys. Rev. Lett.}\ }\textbf {\bibinfo
  {volume} {112}},\ \bibinfo {pages} {044301} (\bibinfo {year}
  {2014})}\BibitemShut {NoStop}%
\bibitem [{\citenamefont {Ott}\ \emph {et~al.}(2020)\citenamefont {Ott},
  \citenamefont {Biehs},\ and\ \citenamefont
  {Ben-Abdallah}}]{PhysRevB.101.241411}%
  \BibitemOpen
  \bibfield  {author} {\bibinfo {author} {\bibfnamefont {A.}~\bibnamefont
  {Ott}}, \bibinfo {author} {\bibfnamefont {S.-A.}\ \bibnamefont {Biehs}}, \
  and\ \bibinfo {author} {\bibfnamefont {P.}~\bibnamefont {Ben-Abdallah}},\
  }\href {\doibase 10.1103/PhysRevB.101.241411} {\bibfield  {journal} {\bibinfo
   {journal} {Phys. Rev. B}\ }\textbf {\bibinfo {volume} {101}},\ \bibinfo
  {pages} {241411} (\bibinfo {year} {2020})}\BibitemShut {NoStop}%
\bibitem [{\citenamefont {Volokitin}(2019)}]{Volokitin2019}%
  \BibitemOpen
  \bibfield  {author} {\bibinfo {author} {\bibfnamefont {A.~I.}\ \bibnamefont
  {Volokitin}},\ }\href {\doibase 10.1134/S002136401911016X} {\bibfield
  {journal} {\bibinfo  {journal} {JETP Letters}\ }\textbf {\bibinfo {volume}
  {109}},\ \bibinfo {pages} {749} (\bibinfo {year} {2019})}\BibitemShut
  {NoStop}%
\bibitem [{\citenamefont {Volokitin}(2020)}]{Volokitin_2020}%
  \BibitemOpen
  \bibfield  {author} {\bibinfo {author} {\bibfnamefont {A.~I.}\ \bibnamefont
  {Volokitin}},\ }\href {\doibase 10.1088/1361-648x/ab71a5} {\bibfield
  {journal} {\bibinfo  {journal} {Journal of Physics: Condensed Matter}\
  }\textbf {\bibinfo {volume} {32}},\ \bibinfo {pages} {215001} (\bibinfo
  {year} {2020})}\BibitemShut {NoStop}%
\bibitem [{\citenamefont {Volokitin}\ and\ \citenamefont
  {Persson}(2020)}]{Volokitin2020}%
  \BibitemOpen
  \bibfield  {author} {\bibinfo {author} {\bibfnamefont {A.~I.}\ \bibnamefont
  {Volokitin}}\ and\ \bibinfo {author} {\bibfnamefont {B.~N.~J.}\ \bibnamefont
  {Persson}},\ }\href {\doibase 10.1088/1361-648x/ab79f8} {\bibfield  {journal}
  {\bibinfo  {journal} {Journal of Physics: Condensed Matter}\ }\textbf
  {\bibinfo {volume} {32}},\ \bibinfo {pages} {255301} (\bibinfo {year}
  {2020})}\BibitemShut {NoStop}%
\bibitem [{\citenamefont {Latella}\ and\ \citenamefont
  {Ben-Abdallah}(2017)}]{PhysRevLett.118.173902}%
  \BibitemOpen
  \bibfield  {author} {\bibinfo {author} {\bibfnamefont {I.}~\bibnamefont
  {Latella}}\ and\ \bibinfo {author} {\bibfnamefont {P.}~\bibnamefont
  {Ben-Abdallah}},\ }\href {\doibase 10.1103/PhysRevLett.118.173902} {\bibfield
   {journal} {\bibinfo  {journal} {Phys. Rev. Lett.}\ }\textbf {\bibinfo
  {volume} {118}},\ \bibinfo {pages} {173902} (\bibinfo {year}
  {2017})}\BibitemShut {NoStop}%
\bibitem [{\citenamefont {Abraham~Ekeroth}\ \emph {et~al.}(2018)\citenamefont
  {Abraham~Ekeroth}, \citenamefont {Ben-Abdallah}, \citenamefont {Cuevas},\
  and\ \citenamefont {García-Martín}}]{doi:10.1021/acsphotonics.7b01223}%
  \BibitemOpen
  \bibfield  {author} {\bibinfo {author} {\bibfnamefont {R.~M.}\ \bibnamefont
  {Abraham~Ekeroth}}, \bibinfo {author} {\bibfnamefont {P.}~\bibnamefont
  {Ben-Abdallah}}, \bibinfo {author} {\bibfnamefont {J.~C.}\ \bibnamefont
  {Cuevas}}, \ and\ \bibinfo {author} {\bibfnamefont {A.}~\bibnamefont
  {García-Martín}},\ }\href {\doibase 10.1021/acsphotonics.7b01223}
  {\bibfield  {journal} {\bibinfo  {journal} {ACS Photonics}\ }\textbf
  {\bibinfo {volume} {5}},\ \bibinfo {pages} {705} (\bibinfo {year} {2018})},\
  \Eprint {http://arxiv.org/abs/https://doi.org/10.1021/acsphotonics.7b01223}
  {https://doi.org/10.1021/acsphotonics.7b01223} \BibitemShut {NoStop}%
\bibitem [{\citenamefont {Ott}\ \emph {et~al.}(2018)\citenamefont {Ott},
  \citenamefont {Ben-Abdallah},\ and\ \citenamefont
  {Biehs}}]{PhysRevB.97.205414}%
  \BibitemOpen
  \bibfield  {author} {\bibinfo {author} {\bibfnamefont {A.}~\bibnamefont
  {Ott}}, \bibinfo {author} {\bibfnamefont {P.}~\bibnamefont {Ben-Abdallah}}, \
  and\ \bibinfo {author} {\bibfnamefont {S.-A.}\ \bibnamefont {Biehs}},\ }\href
  {\doibase 10.1103/PhysRevB.97.205414} {\bibfield  {journal} {\bibinfo
  {journal} {Phys. Rev. B}\ }\textbf {\bibinfo {volume} {97}},\ \bibinfo
  {pages} {205414} (\bibinfo {year} {2018})}\BibitemShut {NoStop}%
\bibitem [{\citenamefont {Moncada-Villa}\ \emph {et~al.}(2015)\citenamefont
  {Moncada-Villa}, \citenamefont {Fern\'andez-Hurtado}, \citenamefont
  {Garc\'{\i}a-Vidal}, \citenamefont {Garc\'{\i}a-Mart\'{\i}n},\ and\
  \citenamefont {Cuevas}}]{PhysRevB.92.125418}%
  \BibitemOpen
  \bibfield  {author} {\bibinfo {author} {\bibfnamefont {E.}~\bibnamefont
  {Moncada-Villa}}, \bibinfo {author} {\bibfnamefont {V.}~\bibnamefont
  {Fern\'andez-Hurtado}}, \bibinfo {author} {\bibfnamefont {F.~J.}\
  \bibnamefont {Garc\'{\i}a-Vidal}}, \bibinfo {author} {\bibfnamefont
  {A.}~\bibnamefont {Garc\'{\i}a-Mart\'{\i}n}}, \ and\ \bibinfo {author}
  {\bibfnamefont {J.~C.}\ \bibnamefont {Cuevas}},\ }\href {\doibase
  10.1103/PhysRevB.92.125418} {\bibfield  {journal} {\bibinfo  {journal} {Phys.
  Rev. B}\ }\textbf {\bibinfo {volume} {92}},\ \bibinfo {pages} {125418}
  (\bibinfo {year} {2015})}\BibitemShut {NoStop}%
\bibitem [{\citenamefont {Joulain}\ \emph {et~al.}(2015)\citenamefont
  {Joulain}, \citenamefont {Ezzahri}, \citenamefont {Drevillon},\ and\
  \citenamefont {Ben-Abdallah}}]{doi:10.1063/1.4916730}%
  \BibitemOpen
  \bibfield  {author} {\bibinfo {author} {\bibfnamefont {K.}~\bibnamefont
  {Joulain}}, \bibinfo {author} {\bibfnamefont {Y.}~\bibnamefont {Ezzahri}},
  \bibinfo {author} {\bibfnamefont {J.}~\bibnamefont {Drevillon}}, \ and\
  \bibinfo {author} {\bibfnamefont {P.}~\bibnamefont {Ben-Abdallah}},\ }\href
  {\doibase 10.1063/1.4916730} {\bibfield  {journal} {\bibinfo  {journal}
  {Applied Physics Letters}\ }\textbf {\bibinfo {volume} {106}},\ \bibinfo
  {pages} {133505} (\bibinfo {year} {2015})},\ \Eprint
  {http://arxiv.org/abs/https://doi.org/10.1063/1.4916730}
  {https://doi.org/10.1063/1.4916730} \BibitemShut {NoStop}%
\bibitem [{\citenamefont {Zhou}\ \emph {et~al.}(2020)\citenamefont {Zhou},
  \citenamefont {Zhang}, \citenamefont {Qu},\ and\ \citenamefont
  {Yi}}]{ZHOU2020106889}%
  \BibitemOpen
  \bibfield  {author} {\bibinfo {author} {\bibfnamefont {C.}~\bibnamefont
  {Zhou}}, \bibinfo {author} {\bibfnamefont {Y.}~\bibnamefont {Zhang}},
  \bibinfo {author} {\bibfnamefont {L.}~\bibnamefont {Qu}}, \ and\ \bibinfo
  {author} {\bibfnamefont {H.-L.}\ \bibnamefont {Yi}},\ }\href {\doibase
  https://doi.org/10.1016/j.jqsrt.2020.106889} {\bibfield  {journal} {\bibinfo
  {journal} {Journal of Quantitative Spectroscopy and Radiative Transfer}\
  }\textbf {\bibinfo {volume} {245}},\ \bibinfo {pages} {106889} (\bibinfo
  {year} {2020})}\BibitemShut {NoStop}%
\bibitem [{\citenamefont {Khandekar}\ \emph {et~al.}(2018)\citenamefont
  {Khandekar}, \citenamefont {Messina},\ and\ \citenamefont
  {Rodriguez}}]{doi:10.1063/1.5018734}%
  \BibitemOpen
  \bibfield  {author} {\bibinfo {author} {\bibfnamefont {C.}~\bibnamefont
  {Khandekar}}, \bibinfo {author} {\bibfnamefont {R.}~\bibnamefont {Messina}},
  \ and\ \bibinfo {author} {\bibfnamefont {A.~W.}\ \bibnamefont {Rodriguez}},\
  }\href {\doibase 10.1063/1.5018734} {\bibfield  {journal} {\bibinfo
  {journal} {AIP Advances}\ }\textbf {\bibinfo {volume} {8}},\ \bibinfo {pages}
  {055029} (\bibinfo {year} {2018})},\ \Eprint
  {http://arxiv.org/abs/https://doi.org/10.1063/1.5018734}
  {https://doi.org/10.1063/1.5018734} \BibitemShut {NoStop}%
\bibitem [{\citenamefont {Latella}\ \emph {et~al.}(2019)\citenamefont
  {Latella}, \citenamefont {Marconot}, \citenamefont {Sylvestre}, \citenamefont
  {Fr\'echette},\ and\ \citenamefont
  {Ben-Abdallah}}]{PhysRevApplied.11.024004}%
  \BibitemOpen
  \bibfield  {author} {\bibinfo {author} {\bibfnamefont {I.}~\bibnamefont
  {Latella}}, \bibinfo {author} {\bibfnamefont {O.}~\bibnamefont {Marconot}},
  \bibinfo {author} {\bibfnamefont {J.}~\bibnamefont {Sylvestre}}, \bibinfo
  {author} {\bibfnamefont {L.~G.}\ \bibnamefont {Fr\'echette}}, \ and\ \bibinfo
  {author} {\bibfnamefont {P.}~\bibnamefont {Ben-Abdallah}},\ }\href {\doibase
  10.1103/PhysRevApplied.11.024004} {\bibfield  {journal} {\bibinfo  {journal}
  {Phys. Rev. Applied}\ }\textbf {\bibinfo {volume} {11}},\ \bibinfo {pages}
  {024004} (\bibinfo {year} {2019})}\BibitemShut {NoStop}%
\bibitem [{\citenamefont {Ben-Abdallah}\ and\ \citenamefont
  {Biehs}(2016)}]{PhysRevB.94.241401}%
  \BibitemOpen
  \bibfield  {author} {\bibinfo {author} {\bibfnamefont {P.}~\bibnamefont
  {Ben-Abdallah}}\ and\ \bibinfo {author} {\bibfnamefont {S.-A.}\ \bibnamefont
  {Biehs}},\ }\href {\doibase 10.1103/PhysRevB.94.241401} {\bibfield  {journal}
  {\bibinfo  {journal} {Phys. Rev. B}\ }\textbf {\bibinfo {volume} {94}},\
  \bibinfo {pages} {241401} (\bibinfo {year} {2016})}\BibitemShut {NoStop}%
\bibitem [{\citenamefont {Kathmann}\ \emph {et~al.}(2020)\citenamefont
  {Kathmann}, \citenamefont {Reina}, \citenamefont {Messina}, \citenamefont
  {Ben-Abdallah},\ and\ \citenamefont {Biehs}}]{kathmann2020scalable}%
  \BibitemOpen
  \bibfield  {author} {\bibinfo {author} {\bibfnamefont {C.}~\bibnamefont
  {Kathmann}}, \bibinfo {author} {\bibfnamefont {M.}~\bibnamefont {Reina}},
  \bibinfo {author} {\bibfnamefont {R.}~\bibnamefont {Messina}}, \bibinfo
  {author} {\bibfnamefont {P.}~\bibnamefont {Ben-Abdallah}}, \ and\ \bibinfo
  {author} {\bibfnamefont {S.-A.}\ \bibnamefont {Biehs}},\ }\href@noop {}
  {\bibfield  {journal} {\bibinfo  {journal} {Scientific reports}\ }\textbf
  {\bibinfo {volume} {10}},\ \bibinfo {pages} {1} (\bibinfo {year}
  {2020})}\BibitemShut {NoStop}%
\bibitem [{\citenamefont {Ott}\ \emph {et~al.}(2019{\natexlab{a}})\citenamefont
  {Ott}, \citenamefont {Messina}, \citenamefont {Ben-Abdallah},\ and\
  \citenamefont {Biehs}}]{10.1117/1.JPE.9.032711}%
  \BibitemOpen
  \bibfield  {author} {\bibinfo {author} {\bibfnamefont {A.}~\bibnamefont
  {Ott}}, \bibinfo {author} {\bibfnamefont {R.}~\bibnamefont {Messina}},
  \bibinfo {author} {\bibfnamefont {P.}~\bibnamefont {Ben-Abdallah}}, \ and\
  \bibinfo {author} {\bibfnamefont {S.-A.}\ \bibnamefont {Biehs}},\ }\href
  {\doibase 10.1117/1.JPE.9.032711} {\bibfield  {journal} {\bibinfo  {journal}
  {Journal of Photonics for Energy}\ }\textbf {\bibinfo {volume} {9}},\
  \bibinfo {pages} {1 } (\bibinfo {year} {2019}{\natexlab{a}})}\BibitemShut
  {NoStop}%
\bibitem [{\citenamefont {Ott}\ \emph {et~al.}(2019{\natexlab{b}})\citenamefont
  {Ott}, \citenamefont {Messina}, \citenamefont {Ben-Abdallah},\ and\
  \citenamefont {Biehs}}]{doi:10.1063/1.5093626}%
  \BibitemOpen
  \bibfield  {author} {\bibinfo {author} {\bibfnamefont {A.}~\bibnamefont
  {Ott}}, \bibinfo {author} {\bibfnamefont {R.}~\bibnamefont {Messina}},
  \bibinfo {author} {\bibfnamefont {P.}~\bibnamefont {Ben-Abdallah}}, \ and\
  \bibinfo {author} {\bibfnamefont {S.-A.}\ \bibnamefont {Biehs}},\ }\href
  {\doibase 10.1063/1.5093626} {\bibfield  {journal} {\bibinfo  {journal}
  {Applied Physics Letters}\ }\textbf {\bibinfo {volume} {114}},\ \bibinfo
  {pages} {163105} (\bibinfo {year} {2019}{\natexlab{b}})},\ \Eprint
  {http://arxiv.org/abs/https://doi.org/10.1063/1.5093626}
  {https://doi.org/10.1063/1.5093626} \BibitemShut {NoStop}%
\bibitem [{\citenamefont {Ben-Abdallah}(2016)}]{PhysRevLett.116.084301}%
  \BibitemOpen
  \bibfield  {author} {\bibinfo {author} {\bibfnamefont {P.}~\bibnamefont
  {Ben-Abdallah}},\ }\href {\doibase 10.1103/PhysRevLett.116.084301} {\bibfield
   {journal} {\bibinfo  {journal} {Phys. Rev. Lett.}\ }\textbf {\bibinfo
  {volume} {116}},\ \bibinfo {pages} {084301} (\bibinfo {year}
  {2016})}\BibitemShut {NoStop}%
\bibitem [{\citenamefont {Latella}\ \emph {et~al.}(2018)\citenamefont
  {Latella}, \citenamefont {Messina}, \citenamefont {Rubi},\ and\ \citenamefont
  {Ben-Abdallah}}]{PhysRevLett.121.023903}%
  \BibitemOpen
  \bibfield  {author} {\bibinfo {author} {\bibfnamefont {I.}~\bibnamefont
  {Latella}}, \bibinfo {author} {\bibfnamefont {R.}~\bibnamefont {Messina}},
  \bibinfo {author} {\bibfnamefont {J.~M.}\ \bibnamefont {Rubi}}, \ and\
  \bibinfo {author} {\bibfnamefont {P.}~\bibnamefont {Ben-Abdallah}},\ }\href
  {\doibase 10.1103/PhysRevLett.121.023903} {\bibfield  {journal} {\bibinfo
  {journal} {Phys. Rev. Lett.}\ }\textbf {\bibinfo {volume} {121}},\ \bibinfo
  {pages} {023903} (\bibinfo {year} {2018})}\BibitemShut {NoStop}%
\bibitem [{\citenamefont {Messina}\ and\ \citenamefont
  {Ben-Abdallah}(2020)}]{PhysRevB.101.165435}%
  \BibitemOpen
  \bibfield  {author} {\bibinfo {author} {\bibfnamefont {R.}~\bibnamefont
  {Messina}}\ and\ \bibinfo {author} {\bibfnamefont {P.}~\bibnamefont
  {Ben-Abdallah}},\ }\href {\doibase 10.1103/PhysRevB.101.165435} {\bibfield
  {journal} {\bibinfo  {journal} {Phys. Rev. B}\ }\textbf {\bibinfo {volume}
  {101}},\ \bibinfo {pages} {165435} (\bibinfo {year} {2020})}\BibitemShut
  {NoStop}%
\bibitem [{\citenamefont {R{\"o}pke}(2013)}]{ropke2013nonequilibrium}%
  \BibitemOpen
  \bibfield  {author} {\bibinfo {author} {\bibfnamefont {G.}~\bibnamefont
  {R{\"o}pke}},\ }\href {https://books.google.com/books?id=K\_QiECntE5kC}
  {\emph {\bibinfo {title} {Nonequilibrium Statistical Physics}}},\ Physics
  textbook\ (\bibinfo  {publisher} {Wiley},\ \bibinfo {year}
  {2013})\BibitemShut {NoStop}%
\end{thebibliography}
\end{document}